\documentclass[twocolumn]{aastex701}

\usepackage{graphicx}
\usepackage{bm}
 \usepackage{amsmath}
 \usepackage{multirow}

\newcommand{\GC}{\mathrm{GC}}

\newcommand{\f}[2]{\frac{#1}{#2}}
\newcommand{\powten}[1]{\times 10^{#1}}

\newcommand{\kpc}{\,\mathrm{kpc}}

\newcommand{\msun}{\,\mathrm{M}_\odot}

\newcommand{\Gyr}{\,\mathrm{Gyr}}
\newcommand{\kms}{\,\mathrm{km\,s^{-1}}}

\newcommand{\sidmthree}{\mathrm{SIDM_{3.5}}}
\newcommand{\sidmseven}{\mathrm{SIDM_7}}
\newcommand{\sidmfourte}{\mathrm{SIDM_{14}}}
\newcommand{\sidmtwe}{\mathrm{SIDM_{21}}}
\newcommand{\sidmten}{\mathrm{SIDM10_{60}}}

\begin{document}

\title{CROCODILE-SIDM: Tidal Formation of Dark Matter-Deficient Galaxies as a Test Case}

\author[0009-0005-8738-8419,gname=Kenji,sname=Kihara]{Kenji Kihara}
\affiliation{Theoretical Astrophysics, Department of Earth and Space Science, The University of Osaka, 1-1 Machikaneyama, Toyonaka, Osaka 560-0043, Japan}
\email[show]{kihara@astro-osaka.jp}

\author[0000-0001-7457-8487,gname=Kentaro,sname=Nagamine]{Kentaro Nagamine}
\affiliation{Theoretical Astrophysics, Department of Earth and Space Science, The University of Osaka, 1-1 Machikaneyama, Toyonaka, Osaka 560-0043, Japan}
\affiliation{Theoretical Joint Research, Forefront Research Center, Graduate School of Science, The University of Osaka, Toyonaka, Osaka 560-0043, Japan}
\affiliation{Kavli IPMU (WPI), UTIAS, The University of Tokyo, Kashiwa, Chiba 277-8583, Japan }
\affiliation{Department of Physics \& Astronomy, University of Nevada, Las Vegas, 4505 S. Maryland Pkwy, Las Vegas, NV 89154-4002, USA}
\affiliation{Nevada Center for Astrophysics, University of Nevada, Las Vegas, 4505 S. Maryland Pkwy, Las Vegas, NV 89154-4002, USA}
\email[show]{kn@astro-osaka.jp}
 
\author[0009-0005-7604-4944]{Lissandre Bourrat}
\affiliation{Theoretical Astrophysics, Department of Earth and Space Science, The University of Osaka, 1-1 Machikaneyama, Toyonaka, Osaka 560-0043, Japan}
 \affiliation{Laboratoire d’Astrophysique, EPFL, Observatoire de Sauverny, 1290 Versoix, Switzerland}
 \email{lissandre.bourrat@epfl.ch}

\author[0000-0001-8404-3507]{Leonard Romano}
 \affiliation{Excellence Cluster ORIGINS, Boltzmannstr. 2, 85748 Garching, Germany}
 \affiliation{European Southern Observatory (ESO), Karl-Schwarzschild-Stra{\ss}e 2, 85748 Garching, Germany}
 \affiliation{Universitäts-Sternwarte, Fakultät für Physik, Ludwig- Maximilians-Universität München, Scheinerstr. 1, D- 81679 München, Ger-
many}
\email{lromano@usm.lmu.de}

\author[0000-0002-5712-6865,gname=Yuri,sname=Oku]{Yuri Oku}
\affiliation{Theoretical Astrophysics, Department of Earth and Space Science, The University of Osaka, 1-1 Machikaneyama, Toyonaka, Osaka 560-0043, Japan}
\email{oku@astro-osaka.jp}

\author[0000-0003-3467-6079,gname=Daisuke,sname=Toyouchi]{Daisuke Toyouchi}
\affiliation{Theoretical Astrophysics, Department of Earth and Space Science, The University of Osaka, 1-1 Machikaneyama, Toyonaka, Osaka 560-0043, Japan}
\email{toyouchi@astro-osaka.jp}

 \shorttitle{CROCODILE-SIDM: Tidal Formation of DMDGs as a Test Case}
 \shortauthors{Kihara, Nagamine, et al.}

\begin{abstract}
We introduce CROCODILE-SIDM, a framework for treating self-interacting dark matter (SIDM) with the $N$-body part of \textsc{GADGET4-Osaka} code, as part of CROCODILE simulation family. 
As a test case, we investigate the impact of SIDM on the tidal formation of dark matter-deficient galaxies (DMDGs) with velocity-dependent cross-section models. We demonstrate that our implementation reproduces the analytic scattering rate in isolated halos. Including dynamical friction self-consistently, we evolve a dwarf satellite with $M_*=2\times10^8\,\mathrm{M}_\odot$ in a $\sim10^{11}\,\mathrm{M}_\odot$ halo on a decaying orbit around a massive host, comparing CDM with four SIDM cross sections for two initial satellite density profiles: a cuspy Navarro--Frenk--White (NFW) profile and a cored Burkert profile. We find that self-interactions primarily regulate the amount of DM retained between pericentric passages, and that the sign of this effect depends on the initial profile: a larger cross section retains more DM for the Burkert initial condition but less DM for the NFW initial condition. We show that this opposing behavior reflects the direction of SIDM heat conduction, which is set by the halo's evolutionary state at infall. Core formation in the cuspy profile assists DM stripping, whereas tidally accelerated gravothermal contraction in the cored profile suppresses tidal mass loss. Consequently, SIDM can either assist or hinder DMDG formation, depending on the satellite's inner structure, making DMDGs a potential probe of SIDM cross section.
\end{abstract}

\section{Introduction}\label{sec:intro}
The CDM model has been remarkably successful in reproducing the large-scale structure of the Universe and a wide range of cosmological observations. Nevertheless, several discrepancies between predictions of collisionless cold dark matter (CDM) and observations have been reported on galactic and sub-galactic scales \citep{Bullock_2017}, including the core--cusp problem \citep{Flores_1994, Moore_1994}, the too-big-to-fail problem \citep{Boylan-Kolchin_2011, Boylan-Kolchin_2012}, and the diversity of dwarf-galaxy rotation curves \citep{Oman_2015}. While baryonic processes such as supernova feedback can alleviate many of these tensions, it remains unclear whether all small-scale phenomena can be explained solely within the standard CDM framework.

Among the proposed solutions, both baryonic feedback \citep{Navarro_1996, Gnedin_2002, Pontzen_2012, Santos-Santos_2020} and alternative dark matter models have received considerable attention. In particular, self-interacting dark matter (SIDM) \citep{Spergel_2000_SIDM} provides a natural mechanism for redistributing energy within dark matter halos via particle scattering, thereby lowering central densities and forming dark matter cores \citep{Firmani_2000, Burkert_2000, Vogelsberger_2012, Rocha_2013}. SIDM has therefore emerged as one of the most extensively studied alternatives to collisionless CDM, and has increasingly been implemented in cosmological simulations \citep{Robertson_2019, Harvey_2019, Correa_2022, Correa_2025, Silverman_2026_merger}.

Recent observations suggest that deviations from CDM predictions may persist
even in systems where baryonic effects are expected to be minimal.
Ultra-faint dwarf galaxies (UFDs), which are strongly dark matter dominated
and contain only a small stellar component, provide an ideal laboratory for
probing the fundamental properties of dark matter, and observational
analyses have reported evidence for cored density distributions in some of
them \citep{Moskowitz_2020, Almeida_2024}, raising the possibility that dark
matter physics itself shapes the internal structure of low-mass galaxies
\citep{Almeida_2025_SIDM}. A similar conclusion has been drawn at much
higher stellar mass for the almost dark galaxy Nube
($M_*\simeq4\powten{8}\msun$), whose exceptionally extended stellar body
($R_e=6.9\kpc$) resists explanation by CDM simulations with baryonic
feedback but is naturally reproduced by a cored halo
\citep{Montes_2024, Almeida_2025_nube}. Such systems are therefore
particularly useful for testing SIDM models.

The impact of SIDM is expected to depend strongly on a system's characteristic velocity scale. Constraints from galaxy clusters generally favor relatively small self-interaction cross sections, whereas dwarf galaxies prefer substantially larger values ($\sigma/m\approx10$--$100\,\mathrm{cm^2\,g^{-1}}$) \citep{Kaplinghat_2016_sidmvariousscales, Ando_2025, Jiang_2026}. In this regime, cores of the observed size form in dwarf halos over a broad range of cross sections, without requiring a finely tuned value \citep{Elbert_2015}. Independent support for large cross sections comes from strong-lensing perturbers. The subhalo detected in the lens SDSS J0946+1006 \citep{Gavazzi_2008, Vegetti_2010} has an inferred central density far in excess of CDM expectations \citep{Minor_2021, Li_2025}, which is naturally explained by gravothermal core collapse in SIDM but requires cross sections at least as large as those favored by dwarf-galaxy measurements. Consequently, dwarf-galaxy-scale halos are among the most promising environments in which SIDM effects may provide additional constraints. Exploring SIDM in low-mass systems is therefore essential for understanding whether dark matter self-interactions leave observable signatures beyond those produced by baryonic feedback alone.

A particularly intriguing class of objects in this context is the population of dark matter-deficient galaxies (DMDGs). The prototype systems NGC 1052-DF2 and NGC 1052-DF4 were reported to possess unusually low dynamical masses compared to their stellar masses \citep{vanDokkum_2018_nature, vanDokkum_2019_DF4}, implying stellar-to-halo mass ratios far above those expected from conventional abundance-matching relations. Since then, additional DMDG candidates have been identified \citep{Guo_2020_furtherDMDGs}, making the formation mechanism of such galaxies an important open problem in galaxy formation theory.

Several scenarios have been proposed to explain the origin of DMDGs. One possibility is the mini-bullet-cluster scenario, in which high-speed collisions between gas-rich dwarf galaxies separate baryons from dark matter and subsequently produce dark matter-deficient remnants \citep{Silk_2019_minibullet, Shin_2020minibullet, Lee_2021_minibullet, vanDokkum_2022_minibulletnature}. An alternative explanation is the tidal stripping scenario, in which a satellite galaxy orbiting a massive host galaxy loses a substantial fraction of its dark matter through tidal heating and stripping \citep{Ogiya_2018}. Controlled numerical experiments demonstrated that satellites with pre-existing dark matter cores can evolve into DMDGs under strong tidal interactions \citep{Ogiya_2021_tidalpotential}. Furthermore, cosmological simulations have also found DMDG analogs formed through tidal processes \citep{Jackson_2021_tidalcosmological, Moreno_2022_cosmological, He_2026}. Tides also leave imprints on the stellar component itself, which provides additional constraints on the tidal origin of DMDGs \citep{Yin_2026}.

A key limitation of some previous tidal studies is the neglect of dynamical friction. Using fully self-consistent $N$-body simulations, \citet{Katayama_2024} showed that dynamical friction modifies satellite orbital evolution, reducing pericentric distances and enhancing tidal stripping relative to simulations with fixed host potentials. Their results suggest that dynamical friction plays an important role in determining whether a satellite can evolve into a DMDG.

While the effects of SIDM on tidal evolution have begun to be explored, most studies have either neglected dynamical friction or focused on systems with substantially lower satellite-to-host mass ratios \citep{Zeng_2022_SIDMevapotidal, Zhang_2025_DF2, Zhang_2025_DF4}. Consequently, the combined influence of SIDM and dynamical friction on DMDG formation remains poorly understood. In particular, it is unclear whether self-interactions can further enhance tidal mass loss in the low-mass dwarf regime where SIDM effects are expected to be strongest.

In this work, we investigate the formation of DMDGs in SIDM models using the CROCODILE-SIDM simulations. 
The CROCODILE simulation suites\footnote{\url{https://sites.google.com/view/crocodilesimulation/}} are performed with \textsc{GADGET4-Osaka} code \citep{Romano_2022a, Romano_2022b, Oku_2022, Oku_2024} for galaxy formation, but here we turn off star formation and feedback mechanisms and use only the $N$-body part of the code. 
Following the framework established by \citet{Katayama_2024}, we focus on a dwarf satellite with $M_*\sim2\powten{8}\msun$ embedded in a $\sim10^{11}\msun$ halo, where baryonic feedback is expected to be weak, and SIDM effects may be more pronounced. Rather than attempting to reproduce the detailed properties of DF2 or DF4 exactly, we compare CDM and SIDM models to quantify how dark matter self-interactions modify the tidal evolution of low-mass satellites and influence the emergence of dark matter-deficient systems.

\section{Implementation of SIDM}\label{sec:SIDMimplementation}
We implemented DM self-interactions in \textsc{GADGET4-Osaka} using the kernel-overlap method developed by \citet{Rocha_2013}, which is derived from the Boltzmann equation. In this approach, each DM particle is assigned an adaptive smoothing length, $h_i$, which is used to estimate the overlap of the kernel functions between neighboring particles. We determine $h_i$ such that each particle has 32 neighbors within a sphere of radius $h_i$, subject to a minimum value of 6\,pc and a maximum value of 10\,kpc.

Because our simulations include high-velocity encounters between DM particles, adopting a constant self-interaction cross section could overestimate the scattering rate in these regimes. Therefore, we also employ a velocity-dependent, angle-averaged cross section $\langle \sigma \rangle$, assuming a Yukawa interaction potential
between DM particles \citep{Feng_2010_prl, Feng_2010_YukawaSIDM, Loeb_2011_YukawaSIDM}: 
\begin{equation}
\langle\sigma\rangle \approx \left\{ \begin{array}{cc} \frac{4\pi}{m_\phi^{2}}
\beta^2 \ln(1+\beta^{-1}), & \beta \lesssim 0.1, \\ \frac{8\pi}{m_\phi^{2}}
\beta^2/(1+1.5\beta^{1.65}), & 0.1 \lesssim \beta \lesssim 10^3, \\
\frac{\pi}{m_\phi^{2}} \left( \ln\beta + 1 - \frac{1}{2} \ln^{-1} \beta
\right)^2, & \beta \gtrsim 10^3,
\end{array} \right.
\label{eq:yukawa_scattering}
\end{equation}
where $\beta \equiv \pi v_{\sigma}^2/v^2=2 \alpha_d m_\phi / (m_\chi v^2)$, $v_\sigma$ is the characteristic velocity at which the momentum-transfer cross section, $\langle\sigma\rangle v$ reaches its maximum value, $v$ is the relative velocity between two particles, $\alpha_d$ is the effective coupling constant, $m_\phi$ is the mediator mass, and $m_\chi$ is the DM particle mass. No single closed-form expression for the Yukawa transfer cross section is valid across the whole parameter space. When the interaction is weak enough for perturbation theory to hold ($\alpha_d m_\chi/m_\phi \lesssim 1$), the differential cross section $d\sigma/d\Omega$ can be evaluated in the Born approximation and integrated with the weight $(1-\cos\theta)$ to give an analytic transfer cross section \citep{Ibe_2010}; this is frequently approximated in the SIDM literature by
\begin{equation}
    \sigma\equiv\int\frac{\mathrm{d}\sigma}{d\Omega}\mathrm{d}\Omega=\frac{4\pi\alpha_d^2}{m_\chi^2\left(m^2_\phi/m_\chi^2+v^2\right)^2}. 
\end{equation}
Our models instead lie in the classical regime, for which
Equation~(\ref{eq:yukawa_scattering}) provides an accurate numerical fit \citep{Khrapak_2003, Khrapak_2004}. 

For the Monte Carlo scattering calculation, the angle-averaged cross section is evaluated from Equation~(\ref{eq:yukawa_scattering}) at the pairwise relative velocity and is denoted $\sigma/m(v_{\mathrm{rel}})$ hereafter; scattering is then treated as isotropic.

Using the velocity-dependent scattering cross section, $\sigma/m(v_\mathrm{rel})$, derived from Equation~(\ref{eq:yukawa_scattering}), the probability that the $i$-th and $j$-th particles scatter during a timestep is given by
\begin{equation}
    P_{ij}=\frac{\sigma}{m}(v_\mathrm{rel})\,v_\mathrm{rel}\,
    \frac{M_i\Delta t_i+M_j\Delta t_j}{2}\,g_{ij},
    \label{eq:P_ij}
\end{equation}
where $M_i$ and $M_j$ are the masses of the two DM particles, $\Delta t_i$
and $\Delta t_j$ are their respective timesteps, and $g_{ij}$ is the kernel overlap factor introduced by \citet{Rocha_2013}, which quantifies the spatial overlap between the smoothing kernels of the two particles. At each timestep, a uniform random number $\xi\in[0,1)$ is drawn for every pair, and the two particles are scattered if $\xi<P_{ij}$.

Our treatment of $g_{ij}$ differs from that of \citet{Rocha_2013}, who
evaluate the overlap integral directly. As shown in
Appendix~\ref{sec:kernel_function}, the convolution of two Gaussian kernels
is itself a Gaussian, so that the overlap reduces to a single kernel
evaluated at the pair separation with an effective smoothing length. We
adopt this closed form,
\begin{equation}
    g_{ij}=W\!\left(\left|\bm{x}_i-\bm{x}_j\right|;
    \sqrt{h_i^2+h_j^2}\right),
    \label{eq:g_ij}
\end{equation}
but evaluate it with the cubic spline kernel \citep{Monaghan_1985_spline_kernel}
that is widely adopted in smoothed particle hydrodynamics,
\begin{equation}
    W(r;h)=\frac{8}{\pi h^3}
    \left\{
    \begin{array}{lll}
        1-6\left(\frac{r}{h}\right)^2+6\left(\frac{r}{h}\right)^3 &\mathrm{if}\, 0\leq\frac{r}{h} \leq\frac{1}{2},\\
        2\left(1-\frac{r}{h}\right)^3 &\mathrm{if}\, \frac{1}{2}<\frac{r}{h}\leq1,\\
         0 &\mathrm{if}\,\frac{r}{h}>1.
    \end{array}
    \right.
    \label{eq:cubic_spline}
\end{equation}
Equation~(\ref{eq:g_ij}) is therefore an approximation for the
spline: it avoids evaluating the overlap integral for every pair at every
timestep, while preserving the compact support that makes the neighbor
search efficient. Appendix~\ref{sec:test} shows that this prescription
reproduces the analytic scattering rate in isolated halos.

To suppress multiple scattering events within a single timestep, we impose a
timestep criterion based on the pairwise scattering rate, 
\begin{equation}
    \Gamma_{ij}= \frac{\sigma}{m}(v_\mathrm{rel})\,v_\mathrm{rel}\,M_i\,g_{ij},
    \label{eq:pair_rate}
\end{equation}
which is the rate at which particle $i$ gets scattered off its neighbor $j$. We require
\begin{equation}
    \Gamma_{ij}\,\Delta t_{i,\mathrm{scat}} \lesssim \kappa
    \label{eq:dt_scat}
\end{equation}
for every neighbor $j$, where $\kappa$ is a dimensionless parameter controlling the maximum scattering probability per timestep.

This is evaluated together with the gravitational timestep criterion,
\begin{equation}
    \Delta t_{i,\mathrm{grav}} = \sqrt{\frac{2\eta h_{i,\mathrm{soft}}}
    {a_\mathrm{grav}}},
    \label{eq:dt_grav}
\end{equation}
where $h_{i,\mathrm{soft}}$ is the gravitational softening length,
$a_\mathrm{grav}$ is the magnitude of the gravitational acceleration, and
$\eta$ is a dimensionless accuracy parameter. Each particle is advanced with
$\Delta t_i=\min(\Delta t_{i,\mathrm{scat}},\,\Delta t_{i,\mathrm{grav}})$, which is also the timestep entering Equation~(\ref{eq:P_ij}).

Our implementation corresponds to the rare-scattering SIDM regime, in which we model self-interactions as discrete large-angle scattering events. Although we employ a velocity-dependent cross section, we compute the scattering probability using an angle-averaged cross section rather than the full differential cross section. A more complete treatment of velocity-dependent SIDM would require sampling the differential cross section, which can lead to frequent small-angle scatterings and substantially increase the computational cost unless an effective drag-force formalism is adopted \citep{Arido_2025_drag_force_imple}. Previous studies have shown that frequent- and rare-scattering schemes can produce distinguishable outcomes in systems with large self-interaction cross sections ($\sigma/m\sim 1\,\mathrm{cm^2\,g^{-1}}$), particularly during cluster mergers \citep{Fischer_2021_fsidm_rsidm_diff}. However, because our simulations primarily probe higher-velocity interactions and moderate effective scattering rates, we adopt the computationally efficient rare-scattering scheme throughout this work.

We adopt $\kappa=1\powten{-2}$ following \citet{Correa_2022}. For the gravitational timestep criterion, we use $\eta=2.5\powten{-2}$. Smaller values are required to accurately resolve deeply gravothermally collapsed halos \citep{Mace_2024_convergencetestsselfinteractingdark, Palubski_2024_eta_kappa, Silverman_2026_merger}, which are beyond the scope of this work.

\section{Simulation Setup}
We perform self-consistent isolated simulations of a host galaxy 
and a satellite galaxy.

\subsection{Host and Satellite Models}
We generate equilibrium galaxy models composed only of $N$-body particles using \textsc{magi} \citep{Miki_2018_MAGI}. Following \citet{Katayama_2024}, we model the host as a pure DM halo, with no stellar component, whose density follows the NFW profile \citep{NFW_1997_profile}:
\begin{equation}
  \rho(r) = \frac{\rho_s}{(r/r_s) \left(1+r/r_s\right)^2},
  \label{eq:NFW_profile}
\end{equation}
where $\rho_s$ and $r_s$ are the scale density and scale radius, respectively. We adopt $(\rho_s,\, r_s) = (4.1\powten{6}\msun\,\mathrm{kpc^{-3}},\, 54\kpc)$ corresponding to $(M_{200},\, c) = (7\powten{12}\msun,\, 4.2)$, the mass--concentration relation expected for the host galaxy of DF2 (NGC 1052) at the estimated infall epoch ($z\sim1.5$) \citep{Ludlow2016, vanDokkum2018b, Fensch2019, Ruiz-Lara2019, Katayama_2024}.

We model the satellite galaxy with DM, stars, and globular clusters (GCs). In SIDM, the inner density profile at the time of infall depends on the halo's scattering history. Here we define core formation as the stage at which an isothermal core has developed and stopped growing. Reaching this state is a collective process that requires $N_c\sim10$ scatterings per particle \citep{Outmezguine_2023, Yang_2024, Engelhardt_2026}. The corresponding timescale is set by the local density $\rho$ and the velocity dispersion $\sigma_v$ of the halo, 
\begin{equation}
\begin{split}
  \tau_\mathrm{core} &\propto \frac{1}{\rho\sigma_v\left(\sigma/m\right)} \\
  &= 4.7\,\mathrm{Gyr}\left(\frac{N_c}{10}\right)
  \left(\frac{\rho}{10^{7}\msun\,\mathrm{kpc^{-3}}}\right)^{-1} \\
  &\times \left(\frac{\sigma_v}{100\kms}\right)^{-1}
  \left(\frac{\sigma/m}{10\,\mathrm{cm^{2}\,g^{-1}}}\right)^{-1}.
\end{split}
\label{eq:sidm_timescale}
\end{equation}

If the halo age is shorter than $\tau_\mathrm{core}$, the halo has not had time to redistribute energy through self-interactions and retains the cuspy NFW profile predicted by collisionless cosmological simulations \citep{NFW_1997_profile}. In contrast, if the halo is old enough, the system develops a central core that is better described by the Burkert profile \citep{Burkert_1995_profile} owing to the self-interactions of DM:
\begin{equation}
    \rho(r) = 
    \f{\rho_b}{\left(1+\f{r}{r_b}\right)\left(1+\f{r^2}{r_b^2}\right)},
    \label{eq:Burkert_profile}
\end{equation}
where $\rho_b$ and $r_b$ are the scale density and scale radius. For $\rho=10^7\msun\,\mathrm{kpc^{-3}}$ and $\sigma_v=100\kms$, $\tau_\mathrm{core}$ exceeds the halo age of $\lesssim4\Gyr$ implied by our assumed infall epoch of $z\sim1.5$ for most of the cross sections considered here. A cuspy profile at infall is therefore a plausible initial condition across our full range of cross sections. Whether a core has already formed, however, depends sensitively on the inner density and velocity dispersion of the progenitor, which are not well constrained. We therefore adopt the two profiles as limiting cases and run the full set of cross sections for each, allowing a direct comparison with \citet{Katayama_2024}.

For the Burkert case, we follow \citet{Katayama_2024} and adopt $(\rho_b,\, r_b) = (5.6\powten{7}\msun\,\mathrm{kpc^{-3}},\, 4.6\kpc)$. For the NFW case, we use the relation between the scale radii of Burkert and equivalent NFW halos derived by \citet{Rocha_2013}, $r_b=0.7\,r_s$, yielding $(r_s,\,\rho_s)=(9.23\,\mathrm{kpc},\,9.29\powten{6}\msun\,\mathrm{kpc}^{-3})$.

The total stellar mass is fixed at $M_* = 2\powten{8}\msun$.
The stellar particles follow the 3D-deprojected S\'{e}rsic profile 
 \citep{Sersic, Sersic2, Sersic3, Sersic4}:
\begin{equation}
    \rho(r) = 
    \rho_s \left(\f{r}{R_e}\right)^{-p_n} \exp\left[-b_n \left(\f{r}{R_e}\right)^{1/n}\right],
\end{equation}
\begin{equation}
    p_n \approx 1 - \f{0.6097}{n} + \f{0.05463}{n^2},
\end{equation}
\begin{equation}
    b_n \approx 2n - \f{1}{3} + \f{4}{405n} + \f{46}{25515n^2}.
\end{equation}
Here, $R_e$ is the projected half-light radius, and $n$ is the S\'{e}rsic index.
We adopt the parameters $(R_e,\, n) = (1.25\kpc,\, 1.0)$, which are representative of $M_*=2\powten{8}\msun$ 
 \citep{MassSize_Wel2014}, although the observed mass--size relation exhibits substantial scatter.

In addition to the diffuse stellar component, we include a GC system in the satellite. The spatial distribution of the GCs in DF2 remains puzzling and may provide new insights into the formation of DMDGs. The GC system of DF2 is unusually extended, with a projected half-number radius of $\sim 3.1\kpc$ \citep{vanDokkum_2018_nature}, considerably larger than expected if the clusters had migrated toward the center through dynamical friction, as invoked in models of nuclear star cluster formation \citep{Tremaine1975, Antonini2013, NSCreview_Neumayer2020}. Because GCs respond to the inner gravitational potential, which SIDM reshapes, we examine how their spatial distribution and kinematics depend on the self-interaction cross section.

We model each of the 20 GCs as a single massive particle rather than as a live, self-gravitating system resolved by multiple particles. \citet{Dutta_Chowdhury2020} showed that this simplification has little effect on their orbital evolution. Following \citet{Katayama_2024}, we sample their initial positions from a S\'{e}rsic profile with $(R_e,\,n)=(2.0\kpc,\,0.5)$, restricted to galactocentric radii of 1--4\,kpc. Each GC is assigned a mass of $m_\GC=5.5\powten{5}\msun$, consistent with the observed GC masses in DF2 \citep{vanDokkum2018b}.

\subsection{Orbital Configuration}
To specify the satellite orbit, we adopt the orbital parameters $(x_c,\, \epsilon)$ defined by \citet{Lacey&Cole1993}:
\begin{eqnarray}
  x_c &\equiv& \frac{r_c(E)}{r_{200,\,\mathrm{host}}}, \\
  \epsilon &\equiv& \frac{L}{L_c(E)}, 
\end{eqnarray}
where $r_c(E)$ is the radius of the circular orbit with orbital energy $E$, $r_{200,\,\mathrm{host}}$ is the virial radius of the host halo, $L$ is the satellite's orbital angular momentum, and $L_c(E)$ is the angular momentum of the circular orbit with energy $E$.
The parameter $\epsilon$ is the orbital circularity, with smaller values corresponding to more radial orbits.
Following \citet{Katayama_2024}, we adopt $(x_c,\, \epsilon) = (0.8, 0.45)$. Compared with the distribution of infall orbital parameters of subhalos measured in the Bolshoi simulation \citep{Klypin_2011_Bolshoi, vandenBosch_2018}, this choice lies at the tightly bound, low-energy end of the $x_c$ distribution, while the circularity is close to its median. \citet{Katayama_2024} demonstrated that these orbital initial conditions lead to substantial tidal stripping and are therefore favorable for forming a DMDG, at least in the CDM scenario. The simulations begin with the satellite at apocenter, with an initial host--satellite separation of 305\,kpc and an initial velocity of $241\kms$. As shown by \citet{Ogiya_2021_tidalpotential} and \citet{Katayama_2024}, both the apocenter and pericenter gradually decrease owing to tidal heating within the satellite and dynamical friction.

\subsection{SIDM Cross Sections}
\begin{figure}[tbp]
    \centering
    \includegraphics[keepaspectratio,width=0.9\columnwidth]{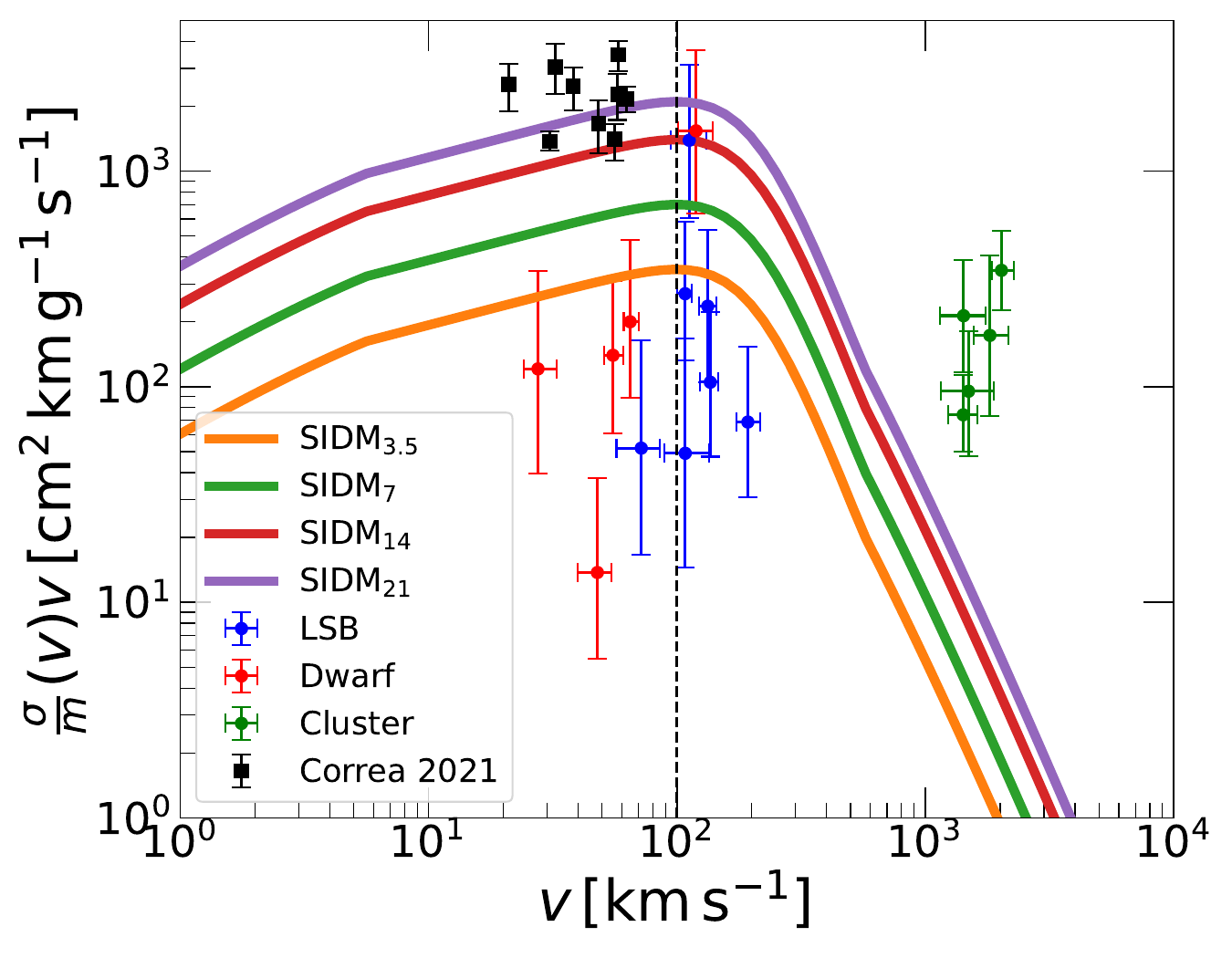}
    \caption{Momentum-transfer cross section of the SIDM models used in this work as a function of velocity. The blue, orange, green, and red solid lines represent $\sidmthree$, $\sidmseven$, $\sidmfourte$, and $\sidmtwe$, respectively. Data points with error bars indicate inferred SIDM cross sections. The colored circles are taken from \citet{Kaplinghat_2016_sidmvariousscales}, where blue, red, and green correspond to low surface brightness (LSB) dwarfs, dwarf galaxies, and galaxy clusters, respectively. The black squares show the values for the classical dwarf spheroidal satellites of the Milky Way from \citet{Correa_2021}. The vertical dashed line marks $v_\sigma=100\kms$.}
    \label{fig:sigmavlog}
\end{figure}

For each initial density profile, we compare simulations with four SIDM cross sections and a CDM model. To isolate the effect of the scattering strength, we fix $v_\sigma=100\kms$ and vary $\sigma/m(v_\sigma)\approx22.7/(m_\phi m_\chi)=3.5$, 7, 14, and $21\,\mathrm{cm^2\,g^{-1}}$. We refer to these models as $\sidmthree$, $\sidmseven$, $\sidmfourte$, and $\sidmtwe$, respectively. Figure~\ref{fig:sigmavlog} shows the momentum-transfer cross section as a function of relative velocity for each SIDM model. The data points indicate cross sections obtained from comparisons between theoretical SIDM halo evolution and observations: \citet{Kaplinghat_2016_sidmvariousscales} estimated the cross
sections required to reproduce the observed inner density profiles of
dwarf galaxies and galaxy clusters, given the age of each system,
while \citet{Correa_2021} modeled the evolution of SIDM halos hosting the
Milky Way's dwarf spheroidal satellites using the gravothermal fluid
formalism combined with orbital evolution and tidal stripping. Our models cover the range of these recent observational estimates.

We note that a recent analysis accounting for gravothermal collapse found a bimodal posterior distribution for the SIDM cross section \citep{Ando_2025} because the stellar kinematics can be explained by either collisionless halos or SIDM halos that have undergone gravothermal collapse, both of which produce centrally concentrated density profiles \citep{Hayashi_2026_constraint}.
We perform a total of ten simulations by combining the two satellite density profiles (NFW and Burkert) with the five dark matter models, as summarized in Table~\ref{tab:simsuite}.
\begin{table}[tbp]
\centering
\caption{Summary of the simulations. IC denotes the initial DM density profile of the satellite, and the cross section is quoted at $v_\sigma=100\kms$.}
\label{tab:simsuite}
\begin{tabular}{lcl}
\hline\hline
IC & $\sigma/m(v_\sigma)$ & Name \\
   & ($\mathrm{cm^{2}\,g^{-1}}$) & \\
\hline
\multirow{5}{*}{Burkert}
 & 0    & BCDM \\
 & 3.5  & B$\sidmthree$ \\
 & 7    & B$\sidmseven$ \\
 & 14   & B$\sidmfourte$ \\
 & 21   & B$\sidmtwe$ \\
\hline
\multirow{5}{*}{NFW}
 & 0    & NCDM \\
 & 3.5  & N$\sidmthree$ \\
 & 7    & N$\sidmseven$ \\
 & 14   & N$\sidmfourte$ \\
 & 21   & N$\sidmtwe$ \\
\hline
\end{tabular}
\end{table}

\subsection{Numerical Parameters}
Because the encounter velocity between the two systems is $\sim700\kms$, far above $v_\sigma$, the cross section for host--satellite pairs is strongly suppressed: for our most extreme model $\sidmtwe$, Equation~(\ref{eq:yukawa_scattering}) gives $\sigma/m\simeq0.13\,\mathrm{cm^{2}\,g^{-1}}$ at this velocity, two orders of magnitude below its value at $v_\sigma$. Evaluating $\Gamma=(\sigma/m)\,\rho_\mathrm{host}\,v_\mathrm{rel}$ at the host density near pericenter, $\rho_\mathrm{host}\sim3\powten{6}\msun\,\mathrm{kpc^{-3}}$, gives $\Gamma\sim0.06\Gyr^{-1}$, so that only of order ten per cent of the satellite particles would evaporate over the full evolution. We have verified this with a lower-resolution run of B$\sidmtwe$ in which host--satellite scattering is enabled: the enclosed DM mass differs from the fiducial run by at most $\sim40\%$ after $10\Gyr$, which is comparable to the uncertainty in the initial conditions and does not affect any of our conclusions. Treating this effect accurately would in any case require equal particle masses in the two systems \citep{Fischer_2021_fsidm_rsidm_diff}, which would substantially increase the computational cost. We therefore turn off scattering between host and satellite DM particles, and also within the host, so that the discussion is confined to the internal response of the satellite. Simulation parameters, including resolution and the number of particles, are summarized in Table~\ref{tab:params}. We adopt a similar resolution to \citet{Katayama_2024}, while increasing the number of GCs to reduce fluctuations in their physical quantities. Gravitational forces are computed with the Tree method \citep{Barnes_1986_TreeMethod}, with an opening-angle criterion of $\theta=0.6$. The centers of the host and the satellite are identified using on-the-fly friends-of-friends (FoF) and the \textsc{subfind} group-and-substructure finder \citep{Springel_2001_subfind1, Springel_2021_subfind2}. 

\begin{table}[tbp]
\centering
\caption{Initial simulation parameters. $N$ and $m$ denote the number and mass of particles of each species, $\epsilon_\mathrm{soft}$ is the gravitational softening length, and the total baryon mass is the sum of the stellar particles and the GCs. $h_\mathrm{si,\,min}$ and $h_\mathrm{si,\,max}$ bound the smoothing length used in the self-interaction calculation, which is set by $N_\mathrm{neighbor}$.}
\label{tab:params}
\begin{tabular}{llc}
\hline\hline
System & Parameter & Value \\
\hline
\multirow{5}{*}{Host halo}
 & $M_{200}$ ($\msun$)                  & $7.00\powten{12}$ \\
 & Total mass ($\msun$)                 & $1.0\powten{13}$ \\
 & $N_\mathrm{DM}$                      & $1.5\powten{7}$ \\
 & $m_\mathrm{DM}$ ($\msun$)            & $6.67\powten{5}$ \\
 & $\epsilon_\mathrm{soft}$ (pc)        & 64 \\
\hline
\multirow{10}{*}{Satellite}
 & Total DM mass ($\msun$)              & $1.00\powten{11}$ \\
 & Total baryon mass ($\msun$)          & $2.11\powten{8}$ \\
 & $N_\mathrm{DM}$                      & $2.5\powten{7}$ \\
 & $N_\star$                            & $5\powten{4}$ \\
 & $m_\mathrm{DM}$ ($\msun$)            & $4.00\powten{3}$ \\
 & $m_\star$ ($\msun$)                  & $4.00\powten{3}$ \\
 & $\epsilon_\mathrm{soft}$ (pc)        & 14 \\
 & $h_\mathrm{si,\,min}$ (pc)           & 6 \\
 & $h_\mathrm{si,\,max}$ (kpc)          & 10 \\
 & $N_\mathrm{neighbor}$                & 32 \\
\hline
\multirow{3}{*}{GCs}
 & $m_\mathrm{GC}$ ($\msun$)            & $5.5\powten{5}$ \\
 & $N_\mathrm{GC}$                      & 20 \\
 & $\epsilon_\mathrm{soft}$ (pc)        & 20 \\
\hline
\end{tabular}
\end{table}

\section{Results}\label{sec:results}
Hereafter, we refer collectively to simulations initialized with Burkert and NFW density profiles as the B-runs and N-runs, respectively, following the naming convention adopted in Table~\ref{tab:simsuite}.

\subsection{Orbital Evolution}
\begin{figure}[tbp]
  \centering \includegraphics[keepaspectratio,width=0.9\columnwidth]{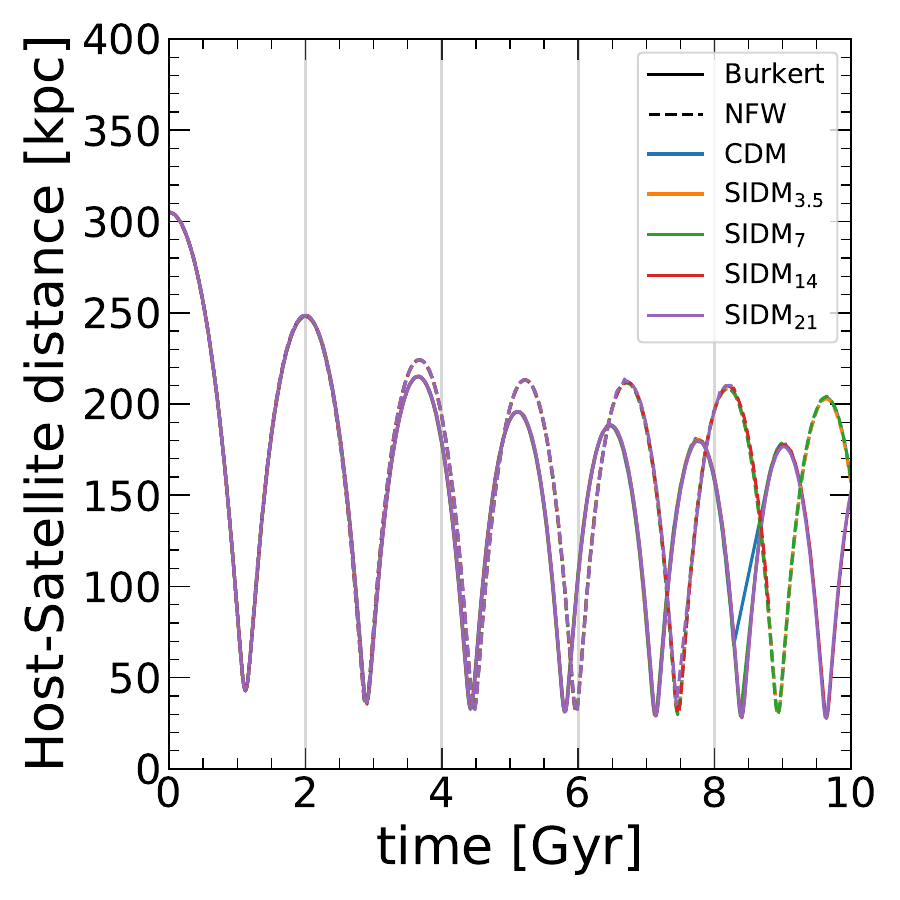}
  \caption{Distance between the centers of the host and the satellite as a function of time. Solid (dashed) lines correspond to Burkert-profile (NFW-profile) initial conditions. Colors indicate different SIDM cross sections: blue, orange, green, red, and purple for CDM, $\sidmthree$, $\sidmseven$, $\sidmfourte$, and $\sidmtwe$, respectively.}
  \label{fig:Orbital_evolution}
\end{figure}

Figure~\ref{fig:Orbital_evolution} shows the evolution of the host--satellite separation over time. For a given initial density profile, varying the SIDM cross section has only a minor impact on the orbital evolution. In contrast, the initial density profile significantly affects the orbital decay. The B-runs undergo seven pericentric passages over the simulation time, whereas the N-runs experience only six. This difference originates from the initial density distributions. Although the Burkert profile has a lower central density, it is denser than the corresponding NFW profile outside $\sim3\kpc$, resulting in higher enclosed mass. Therefore, B-runs experience stronger dynamical friction \citep{Chandrasekhar_1943, Tremaine_1984} and decay more rapidly than in the N-runs. Since the SIDM cross section has little influence on the orbital evolution, we discuss the impact of dark matter self-interactions separately for the B-runs and N-runs in the following sections.

\subsection{Intrinsic Mass Ratio}
\begin{figure}[tbp]
  \centering
 \includegraphics[keepaspectratio,width=0.9\columnwidth]{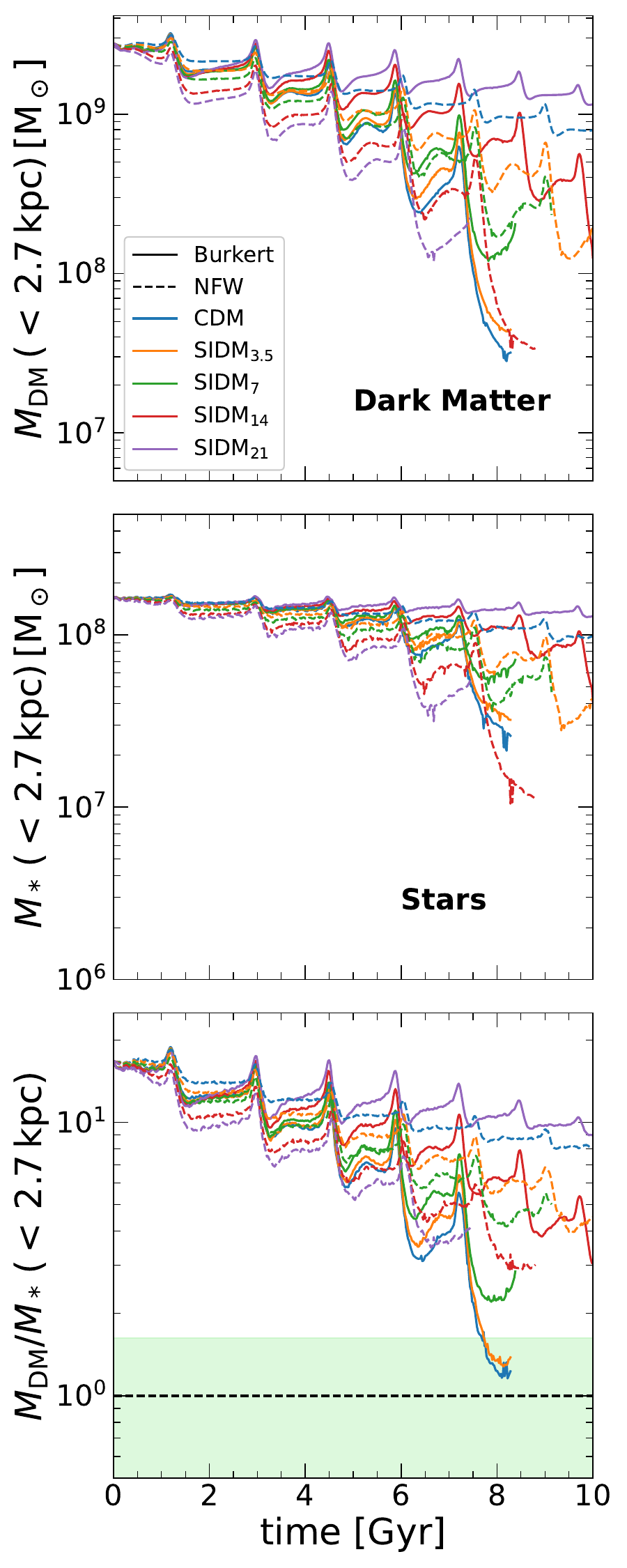}
  \caption{Top panel: DM mass enclosed within $2.7\kpc$ of the satellite center as a function of time. Middle panel: Stellar mass enclosed within $2.7\kpc$ of the satellite center as a function of time. Bottom panel: 3D DM-to-stellar mass ratio, $M_\mathrm{DM}/M_*\,(<2.7\kpc)$, calculated from the top two panels. Line styles and colors are the same as in Figure~\ref{fig:Orbital_evolution}.}
  \label{fig:Enclosed_mass}
\end{figure}

Figure~\ref{fig:Enclosed_mass} shows the evolution of the enclosed DM mass (top), enclosed stellar mass (middle), and the corresponding intrinsic DM-to-stellar mass ratio (bottom), all measured within 2.7\,kpc of the satellite center. This aperture is observationally motivated: adopting the projected circularized half-light radius of DF2, $R_{e,c}=2.0\kpc$ \citep{Cohen_2018}, and the mass estimator
of \citet{Wolf_2010} with $r_{1/2}\simeq(4/3)R_{e,c}$, \citet{Danieli_2019}
derived a dynamical mass of $M_\mathrm{dyn}(<r_{1/2})=(1.3\pm0.8)
\powten{8}\msun$ within $r_{1/2}=2.7\kpc$, comparable to the stellar mass
of $(1.0\pm0.2)\powten{8}\msun$ enclosed by the same radius. This is
therefore the radius at which the dark matter deficiency of DF2 is actually
constrained, and it sets the $M_\mathrm{DM}/M_*\sim1$ criterion adopted
here. We deliberately keep the aperture identical for all runs rather than
rescaling it to the instantaneous, model-dependent half-light radius of
each satellite, so that the mass ratios remain directly comparable both
among the simulations and with the observed value. In runs that undergo severe tidal disruption, the curves truncate after approximately 8\,Gyr because the satellite can no longer be identified by \textsc{subfind}.

The enclosed DM mass exhibits a repeated evolution associated with pericentric passages. It temporarily increases near each pericenter owing to tidal compression and subsequently decreases as tidal stripping removes dark matter. Between successive pericentric passages, the enclosed mass remains nearly constant, consistent with the result of \citet{Katayama_2024}. 

The SIDM cross section primarily affects the amount of DM retained between pericentric passages. In the B-runs, more DM mass is retained as the cross section increases: after the fourth apocentric passage ($t\simeq6.4\Gyr$), B$\sidmtwe$ retains $\sim 5$ times more enclosed DM mass than BCDM. The N-runs show the opposite behavior, with N$\sidmtwe$ retaining $\sim 10$ times \emph{less} than NCDM at the same epoch. The reversal is thus not a marginal effect: it changes the retained mass by nearly an order of magnitude, in opposite directions. The stellar component exhibits the same qualitative dependence on the SIDM cross section, although the variation is less pronounced. Because stars are collisionless, this behavior is induced indirectly through changes in the underlying DM potential. Consequently, the intrinsic DM-to-stellar ratio follows the same trend as the enclosed DM mass. The physical origin of the opposite behavior between the B-runs and N-runs is discussed in Section~\ref{subsec:Discuss_NvsB}.

Based on the 3D mass ratio, only BCDM and B$\sidmthree$ satisfy the criterion for DMDG formation, as shown in the bottom panel of Figure~\ref{fig:Enclosed_mass}. Consistent with \citet{Ogiya_2018}, none of the N-runs, which retain their initially cuspy density profiles, evolve into DMDGs. However, we cannot compare the 3D mass ratio directly with observations because observational estimates rely on projected stellar distributions and line-of-sight kinematics rather than the full 3D mass distribution available in the simulations. To enable a more direct comparison with observations, we construct mock observations of the simulated satellites in the following subsection.

\subsection{Mock-observed Mass Ratio}
To facilitate a more direct comparison with observations, we construct mock observations of the simulated satellites. The simulation coordinate system is defined such that the initial host--satellite separation lies along the $x$-axis, the initial satellite velocity is directed along the $y$-axis, and the resulting orbital angular momentum points along the $z$-axis, which is therefore perpendicular to the orbital plane. We consider three orthogonal lines of sight along the $x$-, $y$-, and $z$-axes. We use these viewing directions consistently throughout the remainder of this section when deriving projected and line-of-sight observables. We denote these lines of sight, as $\mathrm{LOS_x}$, $\mathrm{LOS_y}$, and $\mathrm{LOS_z}$, respectively.

Following \citet{Katayama_2024}, we exclude stellar particles that satisfy either of the following two criteria:~(1) if a stellar particle is the only particle residing in one of $540\,\mathrm{pc}\times 540\,\mathrm{pc}$ pixels in the projected plane, the particle is excluded; and (2) if a particle is more than 10\,kpc away from the satellite center, the particle is excluded. The first criterion is introduced to reproduce the observational limit of the Dragonfly Telephoto Array \citep{Dragonfly}, which observed DF2 \citep{vanDokkum_2018_nature}. Since the mass of the stellar particles is $4000\msun$, the surface brightness of $\mu \lesssim 29\,\mathrm{mag\,arcsec^{-2}}$ corresponds to one particle per $(540\,\mathrm{pc})^2$, assuming that the satellite system is 20\,Mpc away from us and mass-to-light ratio is $M/L_V \sim 3$ from \citet{Bruzual03}. The second criterion ensures that physical values are not contaminated by significantly tidally stripped members.

\begin{figure}[tbp]
    \centering
    \includegraphics[keepaspectratio,width=0.9\columnwidth]{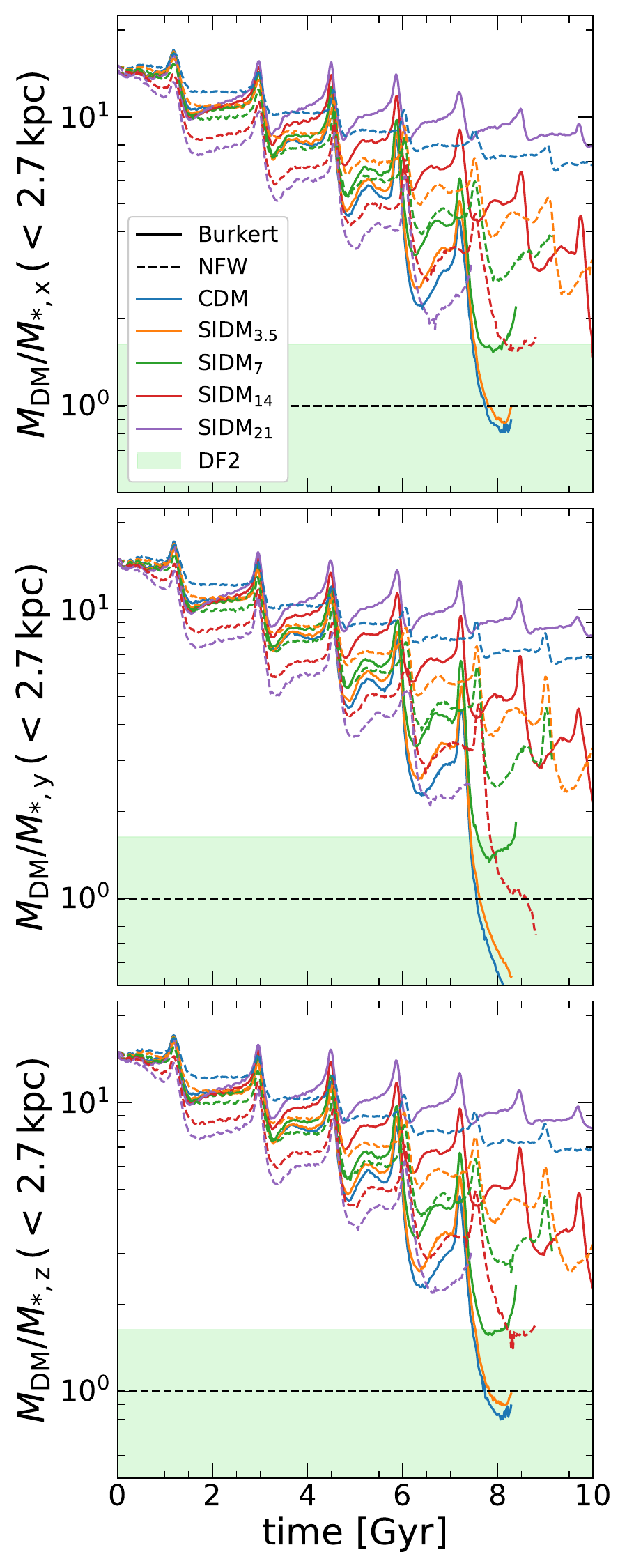}
    \caption{Evolution of the DM-to-stellar mass ratio, $M_\mathrm{DM}/M_\mathrm{*,\,LOS}$, viewed from three different lines of sight. From top to bottom, the panels correspond to the viewing direction of $x$, $y$, and $z$ axes, respectively. Here, $M_\mathrm{*,\,LOS}$ denotes the stellar mass measured within a projected radius of $2.7\kpc$. Line styles and colors are the same as in Figure~\ref{fig:Orbital_evolution}.}
    \label{fig:LOS_mass}
\end{figure}

Figure~\ref{fig:LOS_mass} shows the evolution of the DM-to-stellar mass ratio viewed along the three lines of sight. Only the stellar component is projected here, while the DM mass is kept at its intrinsic three-dimensional value. This is an idealization: in real observations the dynamical mass is inferred from the line-of-sight velocity dispersion, which varies with viewing direction in the same way as the projected stellar mass. A line of sight along which more stars are projected into the aperture also tends to yield a larger $\sigma_\mathrm{LOS}$ and hence a larger inferred dynamical mass, so the mock mass ratio is not necessarily reduced. The values shown here should therefore be read as indicating the amplitude of projection effects rather than as direct predictions for observed mass ratios. A quantitative treatment would require applying the same mass estimator to the mock kinematics, and ultimately a cosmological sample in which the orbits, viewing angles, and assembly histories are not controlled by hand, which we leave to future work. We again recover the trend that a larger cross section yields a higher mass ratio for the B-runs and a lower mass ratio for the N-runs. Notably, however, the number of runs that satisfy the DMDG criterion increases relative to the intrinsic case. B$\sidmseven$ and N$\sidmfourte$ newly exhibit $M_\mathrm{DM}/M_*\sim1$ with a pronounced line-of-sight effect from the $y$-axis. This suggests that, depending on the line-of-sight, we may observe ``apparent'' DMDGs, including intrinsic ones. 

\subsection{Effective Radius and Line-of-sight Velocity Dispersion of the Stellar Component}
\begin{figure*}
    \centering
    \includegraphics[keepaspectratio,width=1.95\columnwidth]{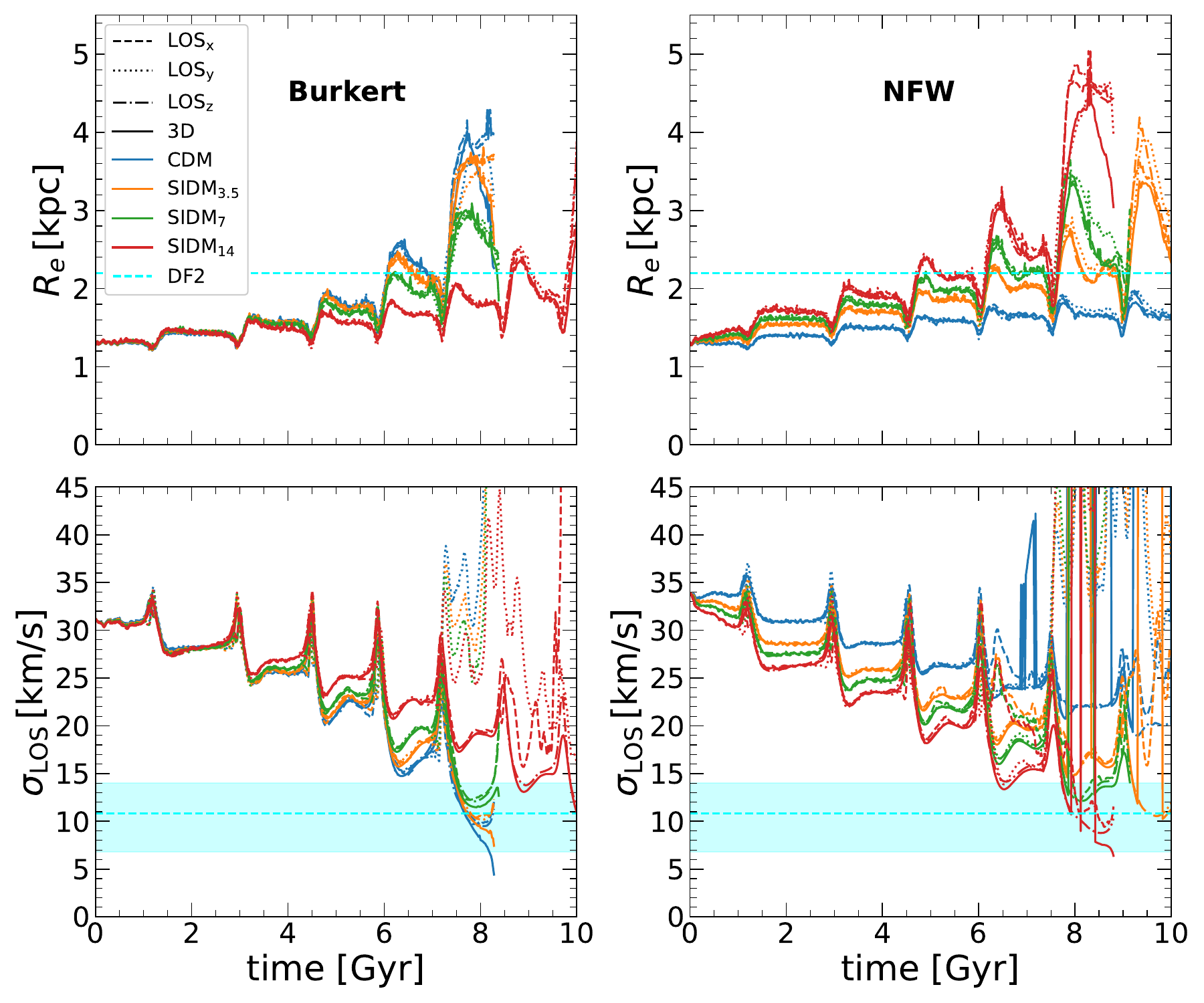}
    \caption{Time evolution of the stellar component of the B-runs (left) and N-runs (right). \textit{Top}:~2D projected half-light radius $R_e$. The cyan horizontal dashed line denotes the measured value of DF2 \citep{vanDokkum2018b}. \textit{Bottom}:~Line-of-sight velocity dispersion $\sigma_\mathrm{LOS}$. The cyan horizontal dashed line and shaded band indicate the measured value and its uncertainty of DF2, respectively \citep{Emsellem2019}. Dashed, dotted, and dashed-dotted lines correspond to values measured from $\mathrm{LOS_x}$, $\mathrm{LOS_y}$, $\mathrm{LOS_z}$, respectively. For comparison, solid lines indicate 3D values assuming spherical symmetry. Colors are the same as in Figure~\ref{fig:Enclosed_mass}.}
    \label{fig:HLR_sigma_star}
\end{figure*}
Figure~\ref{fig:HLR_sigma_star} shows the evolution of the stellar component of the satellite seen from three different lines of sight and measured values, assuming spherical symmetry as well as the observed values of DF2. The upper panels show the time evolution of the 2D-projected half-light radius, $R_e$, while the lower panels show the line-of-sight velocity dispersion, $\sigma_\mathrm{LOS}$. The left (right) column presents the results of the B (N)-runs. We can see that pericentric passages tend to increase $R_e$ and decrease $\sigma_\mathrm{LOS}$, whereas $\mathrm{LOS_y}$ exhibits an opposite trend for $\sigma_{\mathrm{LOS}}$ at later simulation times. This originates from significant tidal disruption at later simulation times.

For the B-runs, as the cross section increases, $R_e$ decreases and $\sigma_\mathrm{LOS}$ increases at the same simulation time. The trend of $R_e$ originates from the difference in the strength of the DM gravitational potential, which retains the stellar component around the central region. For the difference in $\sigma_\mathrm{LOS}$, the virialization of the entire system after pericentric passages affects this trend in the sense that a deeper gravitational potential achieves equilibrium with a higher kinetic energy of the system. These trends lead to a significant difference after 8\,Gyr. While BCDM, B$\sidmthree$, and B$\sidmseven$ exhibit larger $R_e$ than DF2 at 8\,Gyr, B$\sidmfourte$ reaches it after the next pericentric passage. Considering the reproduction of DMDGs at 8\,Gyr in terms of mass ratio, the former three runs fail to reproduce all observables. However, for $\sigma_\mathrm{LOS}$, the former 3 runs align with observation except for $\mathrm{LOS_y}$ while B$\sidmfourte$ exceeds it by a factor of $\sim1.5$ in $\mathrm{LOS_x}$ and $\mathrm{LOS_z}$.

For the N-runs, as the cross section increases, $R_e$ increases and $\sigma_\mathrm{LOS}$ decreases at the same simulation time. This trend is reversed from the B-runs, but the underlying mechanism is identical. While the mass ratio of N$\sidmfourte$ aligns with the DMDG regime at 8\,Gyr, $R_e$ significantly exceeds the observation. However, from $\mathrm{LOS_x}$ and $\mathrm{LOS_z}$, $\sigma_\mathrm{LOS}$ is within the uncertainty of the observation. Notably, N$\sidmseven$ aligns with the observation in terms of both $R_e$ and $\sigma_\mathrm{LOS}$ from $\mathrm{LOS_x}$ and $\mathrm{LOS_z}$ when the mass ratio is $\sim3$, which is almost in the DMDG regime.

\subsection{GC Distribution}

\begin{figure*}
    \centering
    \includegraphics[keepaspectratio,width=1.95\columnwidth]{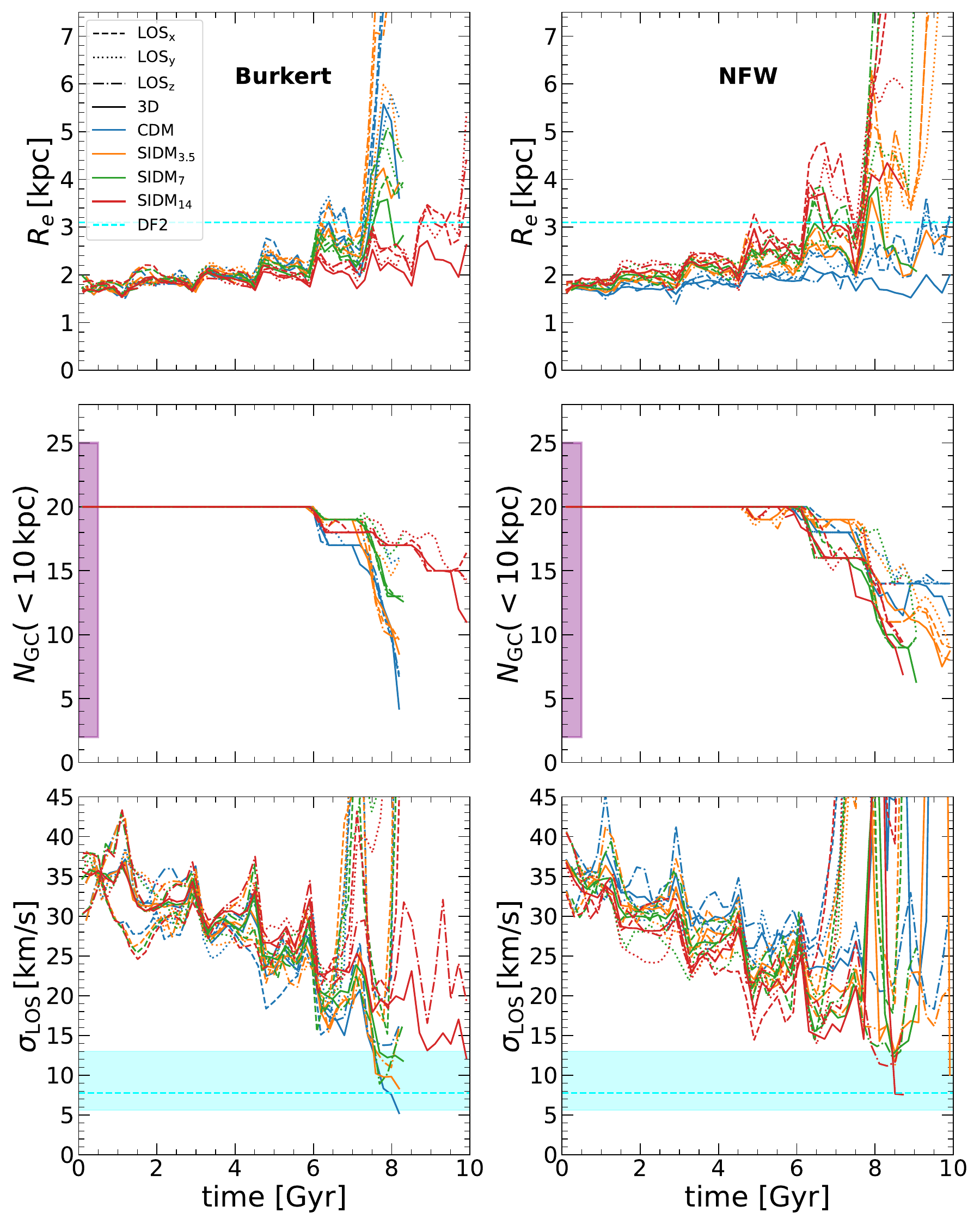}
    \caption{Time evolution of the GCs of the B-runs (left) and N-runs (right). \textit{Top}:~2D projected half-number radius of GCs. The cyan horizontal dashed line denotes the measured value of DF2 \citep{vanDokkum_2018_nature}. \textit{Middle}:~Number of GCs within projected radius of 10\,kpc from the satellite center. Purple hatch on the left edge indicates the observational $1\sigma$ range of low surface brightness galaxies in the Fornax cluster \citep{Prole2019}. \textit{Bottom}:~Line-of-sight velocity dispersion of GCs. The cyan horizontal dashed line and shaded band indicate the measured value and its uncertainty of DF2, respectively \citep{vanDokkum2018_GC}. Line styles and colors are the same as in Figure~\ref{fig:HLR_sigma_star}. We plot averaged values over 200\,Myr because the data points oscillate frequently.}
    \label{fig:GC_props}
\end{figure*}

Figure~\ref{fig:GC_props} shows the evolution of the observable properties of the 20 GCs. Because each run follows only 20 GCs, the GC statistics are inherently noisy; we therefore plot 200\,Myr running averages and focus on robust, systematic trends rather than run-to-run scatter. The top panels plot the 2D projected half-number radius of the GCs. Consistent with the 2D projected half-light radius of the stellar particles, the B-runs exhibit decreasing values with increasing cross section, while the N-runs exhibit the opposite trend. This consistency is a natural consequence of the GCs behaving as collisionless probes that respond only to the gravitational potential modified by SIDM. BCDM, B$\sidmthree$, B$\sidmseven$, and B$\sidmfourte$ align with the observation right after the fourth pericentric passage, while B$\sidmtwe$ aligns with it after the sixth pericentric passage. This means that the B-runs fail to reproduce DM deficiency and $R_e$, simultaneously. Although N$\sidmfourte$ forms a DMDG, its $R_e$ is approximately twice the observed value.

The middle panels plot the evolution of the number of GCs, $N_\mathrm{GC}$, within a projected radius of 10\,kpc from the satellite center. While the B-runs show the first decrease in $N_\mathrm{GC}$ right after the fourth pericentric passage, N$\sidmthree$ and N$\sidmfourte$ show it after the third pericentric passage. We attribute the absence of this feature to its poorer statistics compared with the stellar component, owing to the small number of initial GCs. Similarly to the B-runs, all the N-runs exhibit the onset of a persistent decrease in $N_\mathrm{GC}$ at the fourth pericentric passage. We find that, when the satellite acquires $M_\mathrm{DM}/M_\mathrm{*,\,LOS}\sim1$, BCDM, B$\sidmthree$, B$\sidmseven$, and N$\sidmfourte$ retain 30\%, 50\%, 60\%, and 40\% of the initial number, respectively. Accounting for the tidal loss of GCs, this suggests that the progenitor of DF2 hosted $\sim$20 GCs before its infall into the halo of NGC 1052. The final state of each run retains more than 3 GCs, consistent with the observed range for low-surface-brightness galaxies in the Fornax cluster \citep{Prole2019}.

The bottom panels plot the evolution of the line-of-sight velocity dispersion, $\sigma_\mathrm{LOS}$. The trend for the GCs is again similar to that of the stellar component. Although BCDM, B$\sidmthree$, B$\sidmseven$, and N$\sidmfourte$ exhibit the smallest $M_\mathrm{DM}/M_\mathrm{*,\,LOS}$ along $\mathrm{LOS_y}$, their $\sigma_\mathrm{LOS}$ significantly exceeds the observation. However, the mock observations along $\mathrm{LOS_z}$ reproduce the observation ($\sigma_\mathrm{LOS}\sim 7.8\kms$), except for BCDM, which shows a slight deviation from it.

\section{Discussion}
\begin{figure*}
    \centering
    \includegraphics[keepaspectratio,width=1.95\columnwidth]{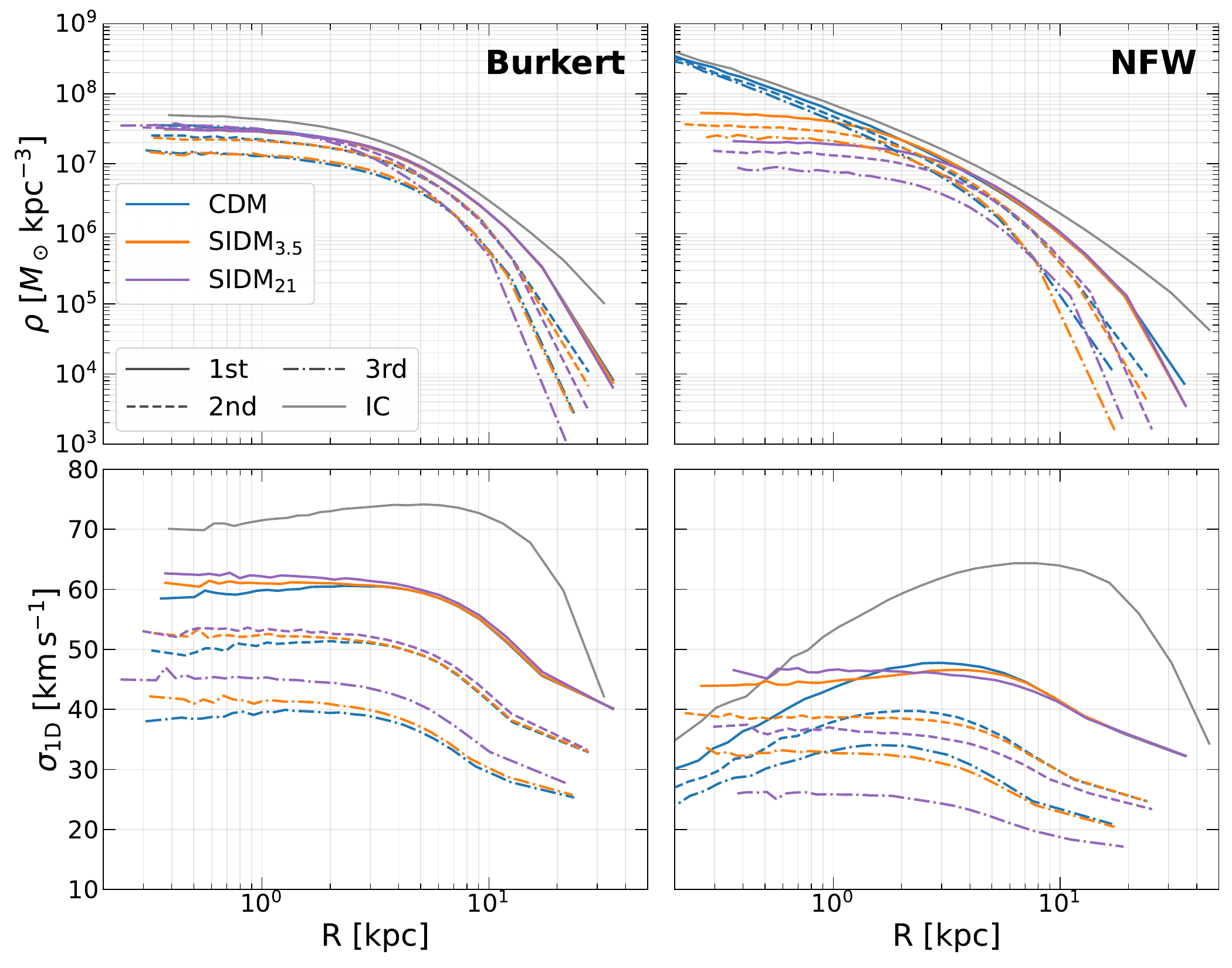}
    \caption{Radial profiles of the subhalo at successive apocentric passages of the B-runs (left) and N-runs (right). \textit{Top}:~dark matter density $\rho$. \textit{Bottom}:~one-dimensional velocity dispersion $\sigma_\mathrm{1D}$. Solid, dashed, and dash-dotted lines correspond to the first, second, and third apocentric passage as indicated in the legend ($t = 2.0$, $3.7$, and $5.1$~Gyr for the B-runs; $t = 2.0$, $3.7$, and $5.2$~Gyr for the N-runs). The grey solid line is the initial condition, which is common to all cross sections. Only the CDM, $\mathrm{SIDM}_{3.5}$, and $\mathrm{SIDM}_{21}$ runs are shown for clarity. Colors are the same as in Figure~\ref{fig:Enclosed_mass}.  
    Corresponding movies are uploaded here: \url{https://www.youtube.com/watch?v=TPGjKzNwx0k} (B-runs) and \url{https://www.youtube.com/watch?v=SUYGXF6-8w4} (N-runs). 
    }
    \label{fig:density_velocity_profiles}
\end{figure*}

The mass-retention trends of Section~\ref{sec:results} reflect how self-interactions reshape the inner structure of the satellite between pericentric passages. Figure~\ref{fig:density_velocity_profiles} shows the DM density and one-dimensional velocity-dispersion profiles measured at the first three apocentric passages, and forms the basis of the discussion below. 
In this figure, we contrast three models: CDM, B$\sidmthree$, and B$\sidmtwe$. 
We first examine the two sets of runs separately, then identify the mechanism responsible for the opposite sign of the SIDM effect.

\subsection{Dynamics in the B-runs} \label{subsec:Discuss_Bruns}
We begin with BCDM in order to separate the effect of tidal interactions from that of self-interactions. The initial profile is flat in density and nearly isothermal in the inner region, with only a shallow positive gradient in $\sigma_\mathrm{1D}$. Tidal interactions preserve the shape of both profiles but lower their normalization at every radius as successive pericentric passages proceed: the central density falls from $\sim3.5\powten{7}$ to $\sim1.5\powten{7}\msun\,\mathrm{kpc^{-3}}$ between the first and third apocentric passages, and $\sigma_\mathrm{1D}$ decreases correspondingly.

The effect of self-interactions is already visible in B$\sidmthree$. By the second apocentric passage, its $\sigma_\mathrm{1D}$ profile is flatter than that of BCDM at the same epoch. Since the two runs differ only in the presence of DM scattering, this flattening must be produced by SIDM thermalization. The density profile, however, is nearly indistinguishable from that of BCDM, indicating that heat redistribution has not yet appreciably altered the mass distribution.

B$\sidmtwe$ shows the effect far more clearly. At each apocentric passage, its innermost $\sigma_\mathrm{1D}$ exceeds that of both BCDM and B$\sidmthree$ and, more strikingly, its central density remains essentially unchanged from the first to the third apocentric passage while that of BCDM drops by a factor of $\sim2$. The satellite has therefore resisted the tidal erosion of its core.

We attribute this to a reversal in the direction of heat conduction. The Burkert initial condition is already close to the isothermal, maximal-core state, so only a modest change in the shape of the $\sigma_\mathrm{1D}$ profile is needed to reverse the heat flux. Tidal stripping supplies exactly that change. Although $\sigma_\mathrm{1D}$ decreases at all radii as mass is removed, it drops disproportionately faster in the outskirts, where the dynamically hotter envelope is stripped first, than in the center. The profile therefore acquires a negative gradient over most of the satellite, and heat is transported outward. Because a self-gravitating system has a negative heat capacity, this outward flux slows the decline of the central temperature and density relative to the CDM run at the same apocentric passage \citep{Colin_2002_sidmhalo}. The magnitude of the flux scales with the cross section, so a larger $\sigma/m$ opposes the tidal reduction of the central density more effectively, which is what produces the increasing enclosed mass seen in Figure~\ref{fig:Enclosed_mass}.

\subsection{Dynamics in the N-runs} \label{subsec:Discuss_Nruns}
In NCDM the cusp stays essentially intact: the density profile remains close to the initial NFW form at all three apocentric passages, with only an overall reduction in normalization. This provides a microscopic view of the result of \citet{Ogiya_2021_tidalpotential}, that cuspy satellites are far more resistant to tidal disruption than cored ones.

Self-interactions change this picture immediately. Already in N$\sidmthree$ the density profile has flattened into a core of $\sim1\kpc$ by the first apocentric passage, and the central $\sigma_\mathrm{1D}$ has risen from $\sim30\kms$ in NCDM to $\sim44\kms$. Both changes follow from the direction of heat conduction: the NFW profile has a strongly inwardly decreasing $\sigma_\mathrm{1D}$, so the center is much colder than its surroundings and heat flows inward, heating and expanding the core. The $\sigma_\mathrm{1D}$ profile is not yet flat at this epoch, indicating that thermalization is still in progress.

N$\sidmtwe$ reaches a lower central density than N$\sidmthree$ at every apocentric passage, and the difference grows with time. The expanded, less tightly bound core is more vulnerable to tidal stripping, and the resulting mass loss lowers both $\rho$ and $\sigma_\mathrm{1D}$ further. Since the scattering rate scales as $\rho\,\sigma_\mathrm{1D}$, this in turn suppresses the rate at which the halo can re-thermalize, so the satellite cannot recover the isothermal configuration that would arrest the expansion. The system is thus locked into a state of progressive expansion and mass loss, which is why a larger cross section yields \emph{less} retained DM.

\subsection{Origin of the Reversed SIDM Effect} \label{subsec:Discuss_NvsB}
Sections~\ref{subsec:Discuss_Bruns} and \ref{subsec:Discuss_Nruns} show that the sign of the SIDM effect on the retained DM mass is set by the direction of heat conduction, which is in turn fixed by the shape of the velocity-dispersion profile at infall. Because tidal mass loss is governed primarily by the mean density interior to the tidal radius, a mechanism that lowers the central density weakens the satellite against stripping, while one that raises it strengthens the satellite. These two initial conditions place the satellite on opposite sides of this divide: the N-runs begin in the core-formation phase, where conduction is inward and the core expands, whereas the B-runs begin near the transition to gravothermal contraction, where conduction reverses and the core recompresses. Tidal interactions reinforce the dichotomy, since removal of the outer, high-entropy envelope shortens the gravothermal timescale and accelerates the onset of core collapse \citep{Nishikawa_2020_tidal, Zeng_2022_SIDMevapotidal}. Our two initial conditions therefore bracket the two qualitatively distinct regimes of tidally evolving SIDM subhalos: monotonic core expansion and tidally accelerated core collapse.

We emphasize that this reversal is not driven by the orbital evolution, which is nearly independent of the cross section at fixed initial profile (Figure~\ref{fig:Orbital_evolution}); it is an internal, structural response of the satellite. A phase-space characterization of the expanding and collapsing cores is left to future work.

\subsection{When Does SIDM Assist or Hinder DMDG Formation?} \label{subsec:Discuss_formation}
The profile dependence identified above implies that self-interactions do not have a monotonic effect on DMDG emergence; they can either promote or inhibit it, depending on the satellite's inner structure at infall. It is therefore essential to ask how natural each initial condition is for a given cross section. Evaluating Equation~(\ref{eq:sidm_timescale}) with $\rho=10^{7}\msun\,\mathrm{kpc^{-3}}$, $\sigma_v=100\kms$ and $N_c=10$ gives $\tau_\mathrm{core}\simeq13$, $6.7$, $3.4$ and $2.2\Gyr$ for $\sidmthree$, $\sidmseven$, $\sidmfourte$ and $\sidmtwe$, respectively, which should be compared with the halo age of $\lesssim4\Gyr$ implied by our assumed infall epoch of $z\sim1.5$. Note that $\tau_\mathrm{core}$ must be evaluated with the density around the scale radius of the progenitor halo before core formation, rather than with that of the resulting core.

Consider first a satellite that falls in with a still-cuspy, NFW-like profile. A larger cross section lowers the central density and makes the halo more vulnerable to tidal stripping, thereby \emph{assisting} DMDG formation. This is consistent with the SIDM studies of DF2 and DF4 by \citet{Zhang_2025_DF2, Zhang_2025_DF4}, who found that SIDM subhalos infalling prior to significant core formation are efficiently stripped into dark matter-deficient remnants. The timescales above show that this channel is internally consistent: for $\sidmthree$ and $\sidmseven$, $\tau_\mathrm{core}$ exceeds the halo age, so the cusp survives to infall, and for $\sidmfourte$, the run that reaches the DMDG regime in our mock observations, $\tau_\mathrm{core}\simeq3.4\Gyr$ is comparable to the halo age, so that a core is only beginning to form. Only $\sidmtwe$ has $\tau_\mathrm{core}$ shorter than the halo age, so a core should have already formed by the infall time. We include its cuspy initial condition not as a realistic configuration but to bound how far SIDM-enhanced stripping can go when the satellite is still cuspy.

Whereas \citet{Zhang_2025_DF2, Zhang_2025_DF4} adopted relatively concentrated stellar distributions, we match the stellar profile to that of \citet{Katayama_2024}, in which the stars are less centrally concentrated relative to the halo. Even in this less favorable configuration, N$\sidmfourte$ is driven into the DMDG regime by SIDM-enhanced stripping (see red dashed line in Figure~\ref{fig:LOS_mass}), showing that SIDM can produce DMDGs from cuspy progenitors across a range of stellar-to-dark-matter concentrations.

The cored channel behaves in the opposite way. When the satellite falls in with a mature core, self-interactions drive the center toward gravothermal contraction and \emph{hinder} DMDG formation, because the recompressed core resists tidal disruption. In our suite, only BCDM and B$\sidmthree$ satisfy the intrinsic DMDG criterion (see blue and orange solid lines in Figure~\ref{fig:Enclosed_mass}), while the stronger cross sections retain progressively more dark matter.

These two runs differ, however, in what they require of the progenitor. BCDM has no self-interactions, so its core must be attributed to baryonic processes, most plausibly repeated supernova-driven outflows \citep{Navarro_1996, Pontzen_2012}, whose efficiency depends on the star formation history and is expected to weaken toward lower stellar masses \citep{Cintio_2014, Tollet_2016, Lazar_2020}. B$\sidmthree$ admits a second possibility, in which the core is generated by the dark matter itself, without fine-tuning of the cross section \citep{Elbert_2015}. It is therefore the only run forming a DMDG in which both the progenitor core and the subsequent tidal stripping can be accounted for within a single dark matter model. What is required here is a core of the observed size at infall, not a fully relaxed one, so the relevant condition is weaker than $\tau_\mathrm{core}$; given that $\tau_\mathrm{core}$ scales inversely with the poorly constrained inner density of the progenitor, the timescales quoted above do not exclude this possibility.

The outcome of tidal DMDG formation is thus highly sensitive to the satellite's gravothermal phase at infall. That phase is not a free parameter but is set by the progenitor's assembly history: \citet{Silverman_2026_merger} showed that mergers inject orbital kinetic energy into a halo and thereby alter its heat transport, so halos with quiescent merger histories proceed to core collapse while those with sustained mergers do not. Which of our two initial conditions a real satellite resembles is therefore determined by its accretion history as much as by the cross section, and cannot be inferred from the halo mass alone. We note that \citet{Silverman_2026_merger} also identify merger-induced heat injection as a route to dark matter-deficient systems in its own right, one that operates before infall and is independent of the tidal channel considered here. Our suite adds two further routes: SIDM-enhanced stripping of a cuspy progenitor at moderate cross sections, and tidal stripping of a pre-existing core that is either baryonic or, for weak self-interactions, of SIDM origin (see Section~\ref{sec:intro} for other scenarios of DMDG formation).

\subsection{Implications for the Observed DMDG Population} \label{subsec:Discuss_observation}
\citet{He_2026} showed that the tidal formation of DMDGs is a common process in the \textsc{colibre} cosmological simulation with CDM. Our results indicate that tidal scenario is sensitive to the self-interaction cross section, and therefore that the abundance of DMDGs may constrain it.

The dependence is in fact monotonic, even though the effect of SIDM reverses with the initial profile. The cross section controls not only the response of the satellite to tides but also its structural state at infall: from Equation~(\ref{eq:sidm_timescale}), $\tau_\mathrm{core}\propto(\sigma/m)^{-1}$, so a larger cross section makes it more likely that a satellite has already thermalized before it enters the host halo. Satellites that infall in this state are protected against tidal disruption by gravothermal contraction, as the B-runs demonstrate. Conversely, a cross section small enough to leave the satellite cuspy at infall is also too small for SIDM-enhanced stripping to be efficient. In both limits, therefore, increasing $\sigma/m$ acts to suppress the tidal formation of DMDGs, and a strongly self-interacting universe should contain fewer DMDGs than a collisionless one.

This offers a route to testing SIDM that complements internal structural measurements. Since the relevant cross section is the one at the velocity scale of the progenitor, and since the surviving stellar component retains a memory of that scale, the observed DMDG population may in principle constrain $\sigma/m(v)$ at dwarf-galaxy velocities. Quantifying this would require a cosmological sample of infalling satellites with a realistic distribution of orbits and gravothermal phases, which is beyond the scope of the controlled experiments presented here.

\section{Conclusion}
We have introduced CROCODILE-SIDM, an implementation of velocity-dependent dark matter self-interactions in the $N$-body part of \textsc{GADGET4-Osaka}, and applied it to the tidal formation of dark matter-deficient galaxies. Building on the framework of \citet{Katayama_2024}, which includes dynamical friction self-consistently, we evolved a dwarf satellite with $M_*=2\powten{8}\msun$ in a $\sim10^{11}\msun$ halo on a decaying orbit around a massive host for $10\Gyr$. For each of two initial satellite profiles, a cuspy NFW and a cored Burkert, we compared CDM with four velocity-dependent SIDM cross sections, $\sigma/m(v_\sigma)=3.5$--$21\,\mathrm{cm^2\,g^{-1}}$ at $v_\sigma=100\kms$, covering the range of recent observations (Figure~\ref{fig:sigmavlog}). Our main findings are as follows.
\begin{enumerate}
\item Self-interactions primarily regulate the amount of dark matter retained between pericentric passages, and the sign of the effect reverses with the initial profile: a larger cross section retains more dark matter for the Burkert initial condition but less for the NFW initial condition, due to the direction of heat conduction (Figure~\ref{fig:density_velocity_profiles}).

\item The stellar and GC observables follow this trend: the half-light and half-number radii decrease with cross section in the B-runs and increase in the N-runs, while $\sigma_\mathrm{LOS}$ behaves oppositely (Figure~\ref{fig:HLR_sigma_star}, \ref{fig:GC_props}).

\item The orbital evolution is nearly insensitive to the SIDM cross section at fixed initial profile, so that the effect of self-interactions is internal to the satellite (Figure~\ref{fig:Orbital_evolution}).

\item SIDM can either assist or hinder DMDG formation. Cuspy satellites infalling before core formation are driven into the DMDG regime by SIDM-enhanced stripping, whereas satellites with a mature core are protected against disruption (Figure~\ref{fig:Enclosed_mass}).

\item Intrinsically, only BCDM and B$\sidmthree$ form DMDGs, but projection effects along favorable lines of sight substantially increase the number of systems that appear dark matter-deficient. N$\sidmseven$ in particular matches both $R_e$ and $\sigma_\mathrm{LOS}$ of DF2 along $\mathrm{LOS_x}$ and $\mathrm{LOS_z}$ while approaching the DMDG regime (Figure~\ref{fig:LOS_mass}).
\end{enumerate}

Taken together, our results show that the impact of SIDM on DMDG formation is not monotonic for individual satellites; rather, it is governed by their gravothermal phase at infall, which itself depends on the SIDM cross section. Because a larger $\sigma/m$ both increases the likelihood that the halo thermalizes before infall and protects an already thermalized satellite, a strongly self-interacting universe should contain fewer tidally formed DMDGs than a collisionless one. The abundance and structural properties of DMDGs may therefore probe dark matter self-interactions at dwarf-galaxy velocities, complementary to constraints from internal density profiles. Quantitative tests of this scenario will require a cosmological sample of infalling satellites spanning realistic distributions of orbits and assembly histories, which we leave to future work.

\begin{acknowledgments}
    Our numerical simulations and analyses were carried out on {\sc SQUID} at the Cybermedia Center, Osaka University as part of the HPCI System Research Project (hp240141, hp250119, hp260040). This work is supported by the MEXT/JSPS KAKENHI Grant Numbers JP22K21349, 24H00002, 24H00241, and 25K01032 (K.N.).
    K.K. was supported by JST SPRING, Grant Number JPMJSP2138.
    K.N. acknowledges the support from the Kavli IPMU, World Premier Research Center Initiative (WPI), UTIAS, the University of Tokyo. L.R. acknowledges the support from the Deutsche Forschungsgemeinschaft (DFG, German Research Foundation) under Germany's Excellence Strategy – EXC 2094/2 – 390783311.
The authors used a large language model to assist with language editing and LaTeX formatting of the manuscript. All scientific content, analysis, and conclusions were developed and verified by the authors, who take full responsibility for the content of this paper.
\end{acknowledgments}

\software{
  Astropy \citep{astropy2013, astropy2018},
  \textsc{gadget-4} \citep{Springel_2021_subfind2},
  \textsc{magi} \citep{Miki_2018_MAGI},
  Matplotlib \citep{Hunter2007},
  Numba \citep{Lam2015},
  NumPy \citep{Oliphant2006, vanderWalt2011, Harris2020},
  Pandas \citep{McKinney2010},
  SciPy \citep{Jones2001, Virtanen2020}
}

\appendix

\section{Computation of Kernel Overlap Factor}\label{sec:kernel_function}
Following \citet{Rocha_2013}, the self-interaction rate is obtained by
coarse-graining the collisional Boltzmann equation with a distribution
function in which each $N$-body particle carries a spatial kernel
$W(|\bm{x}-\bm{x}_p|;h_p)$ and a delta function in velocity. The resulting
pairwise scattering rate between particles $p$ and $q$ is
\begin{equation}
  \Gamma(q|p) = \frac{\sigma}{m}(v_\mathrm{rel})\,M_q\,
  \left|\bm{v}_q-\bm{v}_p\right|\,g_{pq},
  \label{eq:pairwise_rate}
\end{equation}
and the total rate for particle $p$ is $\Gamma(p)=\sum_q\Gamma(q|p)$, which
is symmetrized as $\Gamma_{pq}=[\Gamma(p|q)+\Gamma(q|p)]/2$ to ensure that
each pair is treated consistently. Here $g_{pq}$ is the kernel overlap
factor,
\begin{equation}
 g_{pq} = \int d^3\bm{x}^\prime\,
 W(|\bm{x}^\prime|;h_p)\,W(|\delta\bm{x}_{pq}+\bm{x}^\prime|;h_q),
 \label{eq:gpq}
\end{equation}
where $\delta\bm{x}_{pq}=\bm{x}_p-\bm{x}_q$ is the separation between the two particles. This factor quantifies the spatial overlap of the two smoothing kernels and reduces to a delta function in the point-particle limit. We refer the
reader to \citet{Rocha_2013} for the full derivation of
Equation~(\ref{eq:pairwise_rate}). In the remainder of this appendix, we evaluate
Equation~(\ref{eq:gpq}) analytically for a Gaussian kernel, which motivates the
form implemented in \textsc{GADGET4-Osaka}.

Since the kernels are even functions of $\bm{x}$, the overlap integral in
Equation~(\ref{eq:gpq}) is simply the convolution of the two kernels evaluated at
the pair separation,
\begin{equation}
  g_{pq} = \left(W_{h_p} \ast W_{h_q}\right)\!\left(\delta\bm{x}_{pq}\right),
  \label{eq:gpq_conv}
\end{equation}
where we have introduced the shorthand $W_h(\bm{x}) \equiv W(|\bm{x}|;h)$.
For a Gaussian kernel,
\begin{equation}
  W_h(\bm{x}) = \frac{1}{\left(2\pi h^2\right)^{3/2}}
  \exp\left(-\frac{\bm{x}^2}{2h^2}\right),
  \label{eq:gaussian_kernel}
\end{equation}
the convolution is again a Gaussian whose variance is the sum of the
individual variances, and therefore
\begin{equation}
  g_{pq} = W\!\left(\left|\delta\bm{x}_{pq}\right|;
  \sqrt{h_p^2+h_q^2}\right).
  \label{eq:gpq_gauss}
\end{equation}
In the actual implementation, we approximate the interaction kernel $g_{pq}$ by evaluating Equation~(\ref{eq:gpq_gauss}) with the cubic spline kernel of Equation~(\ref{eq:cubic_spline}) in place of the Gaussian, whereas \citet{Rocha_2013} performed integration of Equation~(\ref{eq:gpq}) directly. Because the spline has compact support, only a finite number of neighbors contribute to the scattering rate of each particle, which substantially simplifies the numerical implementation.

\section{Test Simulations of SIDM}\label{sec:test}
We validate our implementation following \citet{Correa_2022}, using isolated DM halos with a Hernquist profile \citep{Hernquist_1990}. This profile has closed-form expressions for both the density and the radial velocity dispersion,
\begin{align}
    \rho(r) &= \frac{M_{\mathrm{tot}}}{2\pi}\frac{a}{r(r+a)^3}, \label{eq:Hernquist_density}\\
    \sigma^2_{r}(r) &= \frac{GM_{\mathrm{tot}}}{12a} \left\{\frac{12r(r+a)^3}{a^4}\ln\left(\frac{r+a}{r}\right) -\frac{r}{r+a}\left[25+52\frac{r}{a} +42\left(\frac{r}{a}\right)^2 +12\left(\frac{r}{a}\right)^3\right]\right\}, \label{eq:Hernquist_sigma}
\end{align}
where $M_{\mathrm{tot}}$ is the total halo mass and $a$ is the scale radius, allowing a direct comparison between measured and analytic scattering rates in a self-gravitating system. All initial conditions are generated with \textsc{magi} \citep{Miki_2018_MAGI}.

\subsection{Scattering Rate}
Following \citet{Correa_2022}, we first perform runs in which scattering events are identified as in the production runs, but the velocity kicks are not applied. The halo therefore remains in its initial equilibrium, and the measured scattering rate can be compared with the analytic expectation computed from Equations~(\ref{eq:Hernquist_density}) and (\ref{eq:Hernquist_sigma}). The rate is measured by recording the radial position of every pair that satisfies the scattering criterion, $\xi<P_{ij}$, over the first Gyr and binning the events radially.

We test both the constant and the velocity-dependent cross sections. For the former, we adopt $\sigma/m=1\,\mathrm{cm^{2}\,g^{-1}}$ and a halo with $(M_\mathrm{tot},\,a)=(1\powten{14}\msun,\,225\kpc)$; for the latter, we adopt $\sigma/m(v_\sigma=60\kms)=10\,\mathrm{cm^{2}\,g^{-1}}$ (hereafter $\sidmten$) and $(M_\mathrm{tot},\,a)=(1\powten{10}\msun,\,25\kpc)$. To assess numerical convergence, each halo is realized with $64^3$, $128^3$, and $256^3$ particles.

\begin{figure}[tbp]
  \centering
  \includegraphics[keepaspectratio,width=0.5\columnwidth]{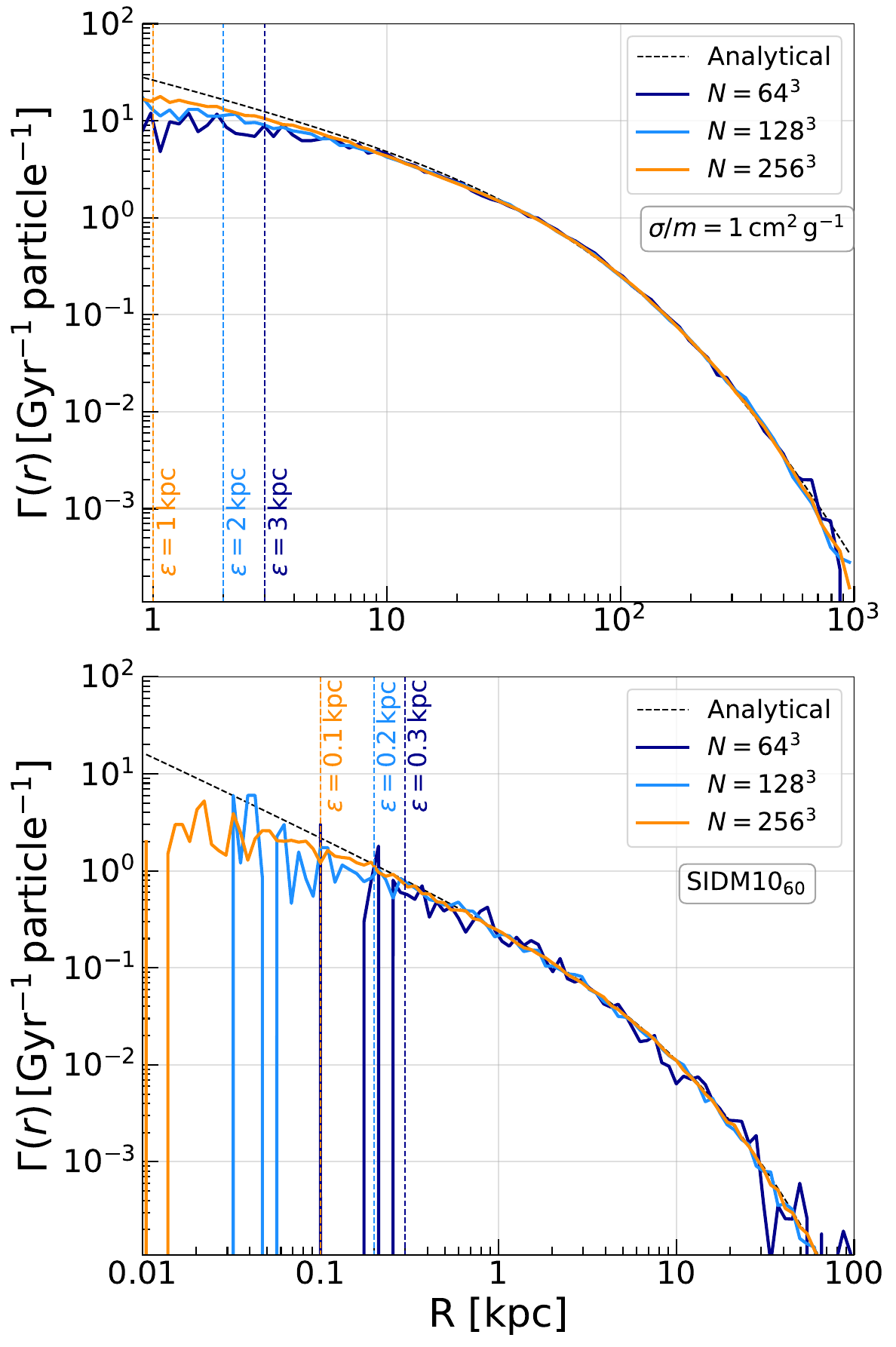}
  \caption{Particle-wise scattering-rate profiles of halos following Equation~(\ref{eq:Hernquist_density}). The top panel shows the scattering rate with a constant cross section, $\sigma/m=1\,\mathrm{cm^{2}\,g^{-1}}$, for a halo initialized with $(M_\mathrm{tot},\,a)=(1\powten{14}\msun,\,225\kpc)$. The bottom panel shows the scattering rate with the velocity-dependent cross section $\sidmten$ for a halo initialized with $(M_\mathrm{tot},\,a)=(1\powten{10}\msun,\,25\kpc)$. The navy, light blue, and orange solid lines indicate runs with $64^3$, $128^3$, and $256^3$ DM particles, respectively, and the black dashed lines show the analytic expectation. Vertical dashed lines mark the softening lengths of the corresponding runs in matching colors.}
  \label{fig:Scattering_profile}
\end{figure}

Figure~\ref{fig:Scattering_profile} shows the resulting particle-wise scattering-rate profiles. Outside the softening length, the measured rates follow the analytic curves at all resolutions, while inside it they are underestimated because gravitational softening lowers the central density. The coarsest runs are noisier but fluctuate around the analytic solution without systematic bias. We adopt softening lengths of $\epsilon=3$, 2, and $1\kpc$ for the $64^3$, $128^3$, and $256^3$ realizations of the $10^{14}\msun$ halo, and $0.3$, $0.2$, and $0.1\kpc$ for the corresponding realizations of the $10^{10}\msun$ halo. These are smaller than the values used by \citet{Correa_2022}, because we find that shallower gravitational potentials reduce the inner DM density and cause significant deviations from the analytic prediction.

\subsection{Evolution of the Test Halo}
\begin{figure}[tbp]
  \centering
  \includegraphics[keepaspectratio,width=\columnwidth]{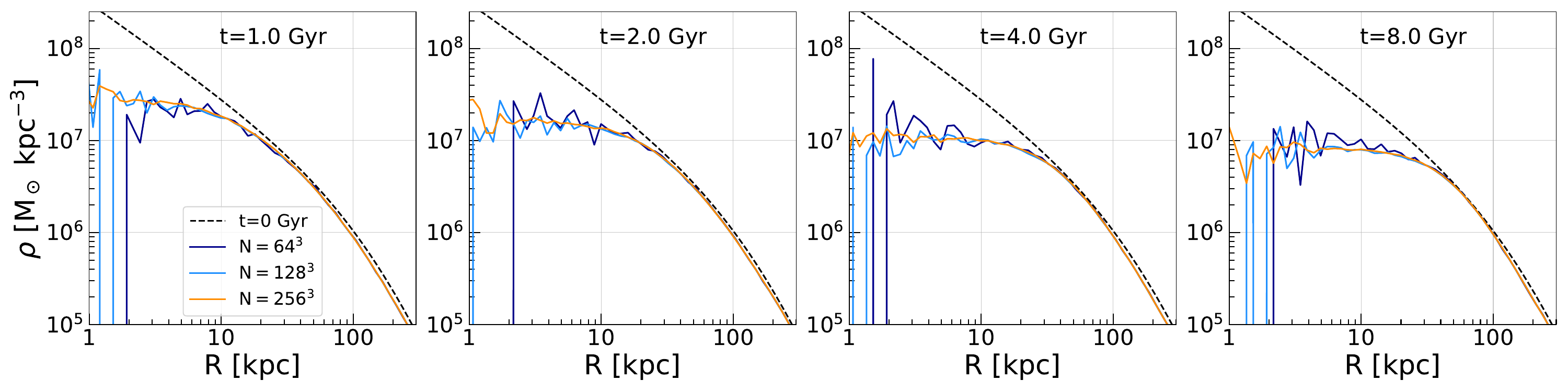}
  \caption{Evolution of the density profile of a halo with $(M_\mathrm{tot},\,a)=(1\powten{14}\msun,\,225\kpc)$ and a constant cross section, $\sigma/m=1\,\mathrm{cm^{2}\,g^{-1}}$, in simulations with different resolutions. From left to right, the panels show snapshots at $t=1$, 2, 4, and 8\,Gyr. The navy, light blue, and orange solid lines indicate runs with $64^3$, $128^3$, and $256^3$ particles, respectively, and the black dashed line shows the initial ($t=0$) profile.}
  \label{fig:Density_evolution_resolution}
\end{figure}

We next enable the velocity kicks and follow the gravothermal evolution of the halo with $(M_\mathrm{tot},\,a)=(1\powten{14}\msun,\,225\kpc)$ and $\sigma/m=1\,\mathrm{cm^{2}\,g^{-1}}$. Figure~\ref{fig:Density_evolution_resolution} shows the density profiles at $t=1$, 2, 4, and 8\,Gyr for the three resolutions. A constant-density core forms and grows with time, and the results converge with resolution: the coarsest run fluctuates around the two finer runs without systematic offset.

\begin{figure}[tbp]
  \centering
  \includegraphics[keepaspectratio,width=0.9\columnwidth]{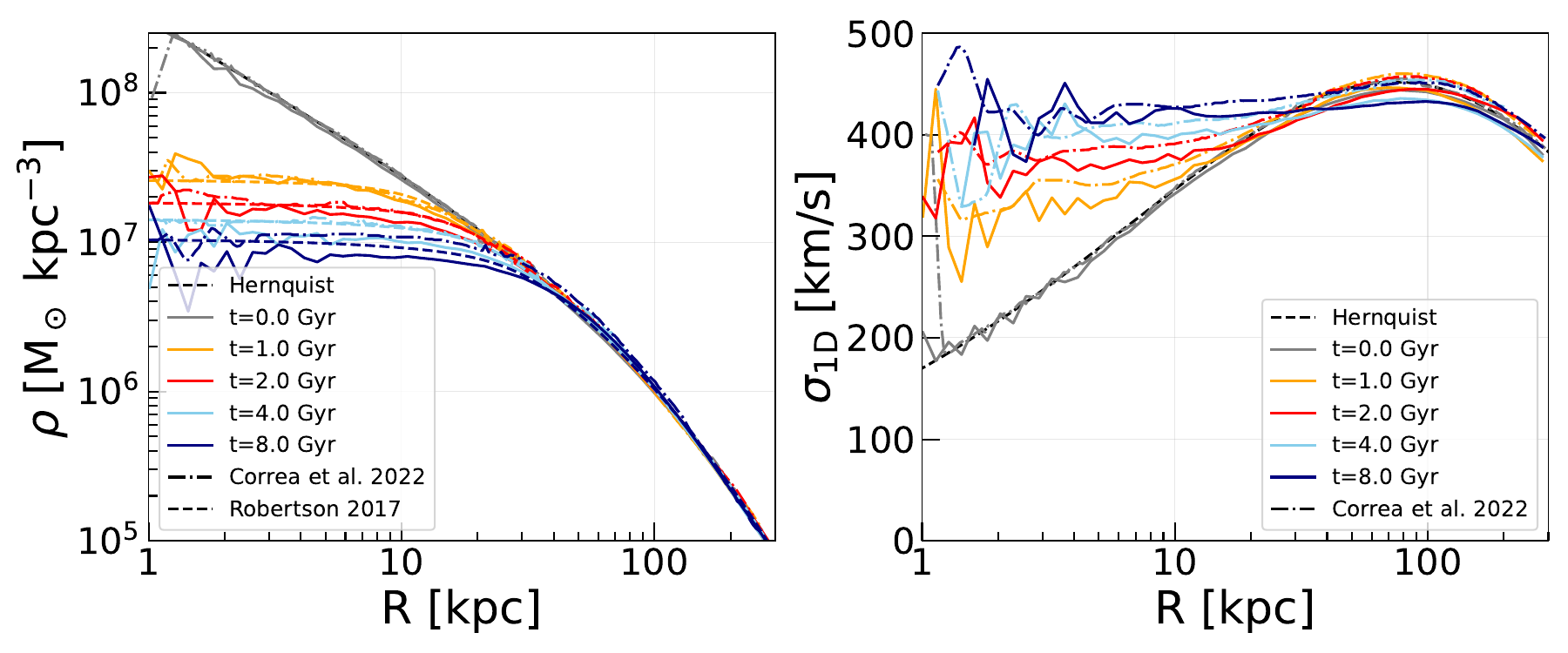}
  \caption{Evolution of the density (left panel) and one-dimensional velocity dispersion (right panel) profiles of the same halo as in Figure~\ref{fig:Density_evolution_resolution}, realized with $256^3$ particles. Note that colors here denote epochs rather than resolutions: the orange, red, sky blue, and navy lines represent snapshots at $t=1$, 2, 4, and 8\,Gyr, respectively, and the gray lines show the initial condition. Solid lines show our results, while dashed-dotted and dashed lines show the previous results of \citet{Correa_2022} and \citet{Robertson_2017}, respectively. Black dashed lines indicate the analytic profiles of Equations~(\ref{eq:Hernquist_density}) and (\ref{eq:Hernquist_sigma}).}
  \label{fig:Profile_evolution_previous}
\end{figure}

Figure~\ref{fig:Profile_evolution_previous} compares the $256^3$ run with the simulations of \citet{Robertson_2017} and \citet{Correa_2022}. Our run reproduces the expected gravothermal behavior: the central density decreases as the core grows, while the inner velocity-dispersion profile flattens and rises toward isothermality. Quantitatively, our initial velocity-dispersion profile deviates slightly from theirs in the innermost region, and this small initial offset propagates into a marginally lower central density and velocity dispersion at late times. We attribute this residual to differences among initial-condition generators: our equilibrium halos are constructed with \textsc{magi}, whereas the generators used in the reference simulations are not documented, so the offset cannot be reduced further without access to identical initial conditions. Since the purpose of this appendix is to verify that our implementation captures the known gravothermal evolution rather than to reproduce a specific realization, we regard the level of agreement in Figure~\ref{fig:Profile_evolution_previous} as sufficient for the applications in this paper.

\bibliography{mybib}
\bibliographystyle{aasjournalv7}
\end{document}